\documentclass[a4paper, usenatbib]{mnras}

\usepackage[T1]{fontenc}
\usepackage{graphicx}	
\usepackage{amsmath}	
\usepackage{amssymb}	
\usepackage{comment}
\usepackage{multirow}
\usepackage{xcolor}
\usepackage{float} 

\title[Spectral-Recurrence of Cyg X-1]{Correlated Spectral and Recurrence Variations of Cygnus X-1}

\author[Broadbent \& Phillipson]{
E. M. Broadbent\href{https://orcid.org/0000-0001-9225-4136}{\textcolor[HTML]{A6CE39}{\includegraphics[scale=0.5]{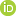}}}$^{1}$
and R.\,A.~Phillipson\href{https://orcid.org/0000-0001-6891-7091}{\textcolor[HTML]{A6CE39}{\includegraphics[scale=0.5]{ORCID_icon.png}}}$^{2}$\thanks{Corresponding Author, Email: rebecca.phillipson@villanova.edu}
\\
$^{1}$DiRAC Institute, Department of Astronomy, University of Washington, 3910 15th Avenue NE, Seattle, WA 98195, USA\\
$^{2}$Villanova University, Department of Physics, Villanova, PA 19085, USA
}

\date{Accepted 2023 November 22. Received 2023 November 22; in original form 2022 November 09\\
\textit{This article has been accepted for publication in Monthly Notices of the Royal Astronomical Society Published by Oxford University Press on behalf of the Royal Astronomical Society.}}

\pubyear{2023}

\begin{document}
\label{firstpage}
\pagerange{\pageref{firstpage}--\pageref{lastpage}}
\maketitle

\begin{abstract}

We present results of recurrence analysis of the black hole X-ray binary Cygnus X-1 using combined observations from the Rossi X-ray Timing Explorer All-sky Monitor and the Japanese Monitor of All-sky X-ray Image aboard the ISS. From the time-dependent windowed recurrence plot (RP), we compute ten recurrence quantities that describe the dynamical behavior of the source and compare them to the spectral state at each point in time. We identify epochs of state changes corresponding to transitions into highly deterministic or highly stochastic dynamical regimes and their correlation to specific spectral states. We compare k-Nearest Neighbors and Random Forest models for various sizes of the time-dependent RP. The spectral state in Cygnus X-1 can be predicted with greater than 95 per cent accuracy for both types of models explored across a range of RP sizes based solely on the recurrence properties. The primary features from the RP that distinguish between spectral states are the determinism, Shannon entropy, and average line length, all of which are systematically higher in the hard state compared to the soft state. Our results suggest that the hard and soft states of Cygnus X-1 exhibit distinct dynamical variability and the time domain alone can be used for spectral state classification.

\end{abstract}




\section{Introduction}

Accreting black holes and neutron stars in X-ray binaries (XRBs) are particularly luminous in the X-rays. The X-ray radiation produced by accretion leads to complex phenomenology in both the energy spectra and timing properties of these exotic systems. In energy spectra, many black hole XRBs transition between a lower luminosity state with a spectrum consisting of harder X-rays (the ‘low-hard’ state, or LHS) and a higher luminosity state with a spectrum consisting predominantly of softer X-rays (the ‘high-soft’ state, or HSS; e.g., \citealt{Homan2001}). The widely used `truncated disk/hot inner flow' model \citep{Ichimaru1977, Narayan1995, Narayan2008} determines that these two states correspond to emission dominated by thermal Comptonization in a hot, geometrically thick, and optically thin plasma in the LHS \citep{Haardt1993, Dove1997, Zdziarski2003} and emission dominated by a geometrically thin and optically thick standard accretion disk \citep{Shakura1973} emulating a pseudo-blackbody spectrum in the HSS (\citealt{Gierlinski1999}, \citealt{Dotani1997}; see review by \citealt{done2007} and references within). 

Cygnus X-1 canonically demonstrates the transitions between the HSS and LHS in its X/$\gamma$-ray emission (e.g., \citealt{Gierlinski1999}). It is supposed that the transition between these two states involves the variation in the location of the inner edge of the thin accretion disk – in the LHS the disk is truncated and instead filled with a hot, optically thin corona (e.g., an advection-dominated accretion flow; \citealt{Narayan1995}) and in the HSS, the cooler disk extends closer to the innermost stable circular orbit about the black hole (e.g., \citealt{Gierlinski2002}, \citealt{Steiner2010}). Although models of the HSS are consistent with a dominating optically thick, geometrically thin accretion disk, the cause and location of a receding accretion disk remain unclear (e.g. \citealt{Miller2006, Petrucci2014, Done2010}). Some theories associate the coronal emission that produces the non-thermal, power-law components in the energy spectra to the base of a jet \citep{Markoff2005}. Indeed, observations of time lags between the hard X-ray corona and reflected emission from the softer X-ray disk (referred to as `reverberation lags'; \citealt{Uttley2014}) enable a full mapping of the disk-corona geometry. As the system evolves between the soft and hard spectral states, \cite{Wang2022} posits that the inner edge of the thin accretion disk recedes and is replaced by a contracting corona. \cite{Wang2022} suggest a model in which emission from the disk and corona are connected and delayed: seed photons from the disk (which additionally undergoes driving mass accretion fluctuations that propagate) are provided to the corona, where coronal heating occurs. The relative accretion fluctuation and coronal heating evolve as the inner edge of the disk recedes/fills in and the corona contracts/expands. 

Other theories of what causes state changes are based on a two accretion flow model, such as that presented by \cite{Smith2002} of a Keplerian accretion disk in combination with a hot, sub-Keplerian halo, as originally posited by \cite{Chakrabarti1995}. A boost in the accretion rate will occur on different timescales in the disk (viscous timescale) versus the halo (close to free fall timescale), which can explain the delayed response of the soft component often observed. \cite{Soria2011} refined this idea: the accretion flow switches between a hard component sourced by a magnetically powered coronal outflow and the soft accretion disk because of changes in the poloidal magnetic field. It is clear the mechanisms responsible for state changes are still debated, involving multiple components of the accretion environment.

XRBs also exhibit complex temporal variability across many different timescales, ranging from sub-second Quasi-periodic Oscillations (QPOs; e.g., see review by \citealt{Ingram2019} and references within) to super-orbital modulations on the order of hundreds of days (e.g., \citealt{Sood2007}). It is hypothesized that there may be a single origin for most of the timing variability features in the power spectra of accreting black holes and neutron stars \citep{Maccarone2011}, where multiple frequencies tend to be well correlated with one another and with the different spectral states of the source \citep{Psaltis1999, Wijnands1999}. The timing variations of XRB light curves presumably carry information about the spectral changes and thus nature and evolution of the accretion history. Characterizations of the temporal behavior of XRBs and the connection to the accretion flow are important for determining universal models of accretion as a function of mass (e.g. see \citealt{Scaringi2015} connecting Young Stellar Objects, White Dwarfs, XRBs, and AGN by their timing behavior) and for classifying objects in large data sets for which spectral information is limited or not attainable.

There have been a wide variety of studies of the complex timing behavior in XRBs revealing intrinsic variability beyond those possible using power spectrum techniques alone, suggesting methods based on second-order moments are insufficient for capturing behavior that is possibly nonlinear or non-stationary. For example, the non-Gaussian and non-zero skewness values of the temporal variation of Cygnus X-1 suggested that the variations are nonlinear in nature \citep{Thiel2001, Timmer2000, Maccarone2002}. \cite{Misra2004} similarly found that the temporal behavior of the black hole system GRS 1915+105 is governed by a low-dimensional chaotic system, detectable when the variability rises above the Poisson fluctuations. In a previous study, \cite{Phillipson2018} confirmed chaotic behavior in the long-term light curve of a neutron star XRB with a superorbital period of approximately 120 days. Finally, using similar nonlinear time series analysis techniques, \cite{Sukova2016} applied methods from recurrence analysis to distinguish stochastic and chaotic rapid variability classes among six microquasars. 

There are abundant spectral-timing studies of black hole XRBs in the fast variability domain (for example, connecting the appearance of QPOs to the hard spectral states; \citealt{Remillard2006}). One of the outstanding questions in the study of the timing variability of accreting sources is to what extent the complex, and possibly nonlinear or aperiodic, variability is connected to the geometry of the accretion flow \citep{BoydSmale2004}. More precisely, there is an ongoing effort to determine whether the structure of the temporal flux variability can elucidate the spectral state of the accreting source. Such efforts are readily supported by advances in machine learning. For example, \cite{Huppenkothen2017} used summary statistics of the raw light curve, power spectrum features, and hardness ratios to build a machine learning model that classified the canonically chaotic black hole XRB, GRS 1915+105, into its variability classes (e.g., `$\rho$,' `$\theta$,' `$\kappa$,' etc. variability states as defined by \citealt{Belloni2000}, \citealt{KleinWolt2002}, and \citealt{Hannikainen2003}). A similar goal was implemented by \cite{OrwatKapola2022}, demonstrating how a neural network and Gaussian mixture model can be used to predict the variability classification of GRS 1915+105. Both these studies demonstrate the complexity of temporal variability and infer its relationship to spectral behavior, since the models depend on the spectral information of the source. However, the models predicted the variability classification type, rather than the accretion state. The recent study by \cite{Sreehari2021} showed that a machine-learning model could be used to predict the spectral states of outbursting black hole XRBs. However, the model they use is based on features that include spectral information: X-ray flux, hardness ratios, presence of various types of QPOs, photon indices, and disc temperature. 

In this study, we aim to build on the recent success in applying machine learning to pattern recognition of variability patterns in XRBs to predict the spectral state of Cygnus X-1 specifically based only on temporal variability patterns (i.e., where the spectral information does not factor into the predictor variables of the model).  

Cygnus X-1 (hereafter, Cyg X-1) is the archetypical black hole XRB and has been the subject of a long history of observational and theoretical studies of accretion onto compact objects since its discovery in 1964 \citep{Bowyer1965}. Cyg X-1 consists of a black hole with a historically measured mass of around 10-15 solar masses, with a recent measurement of up to 21 solar masses \citep{miller-jones2021} at approximately 2 kpc away \citep{reid2011}. Based on the recent mass measurement of approximately 21 solar masses, the spin parameter of the black hole has been constrained to be extreme, at a* > 0.998 \citep{Zhao2021, Kushwaha2021}.

Cyg X-1 is a persistent emitting source in the X-ray, where the primary source of accretion is via stellar wind from an OB supergiant \citep{Bolton1972, Walborn1973}. Most relevant to our study, Cyg X-1 transitions between the LHS and HSS regularly. In the LHS, Cyg X-1 contains persistent radio emission from an unresolved core and a variable relativistic jet \citep{Stirling2001}. Recently, long-term radio/X-ray monitoring of Cyg X-1 has revealed that a compact radio jet also persists in the soft spectral state \citep{Zdziarski2020}, where the jet radio emission directly correlates with the hard X-rays (above 15 keV) and lags the soft X-rays by 100 days. The implication is that the disk corona and jet are powered by the same physical process and the hard X-rays in particular are a direct probe of these correlations (while the disk component potentially varies independently).

In this study, two machine learning techniques are used to find a correlation between variability features evident in the long-term light curve of Cyg X-1 and its known spectral behavior. Specifically, we explore how temporal variability classes described by recurrences in phase space correlate with the high and low spectral states defined by the hardness intensity diagram. The recurrent behavior of a system as it evolves in a phase space picture is distinct between stochastic, linear, periodic, and nonlinear dynamical systems and it is possible to connect temporal dynamical regimes manifesting in the light curve to specific accretion states. Furthermore, recurrence analysis is useful for detecting dynamical regime changes, e.g., from quasi-periodicity\footnote{We note that the term `quasi-periodicity' used here refers to the definition from nonlinear dynamics: the occurrence of multiple unstable periodic orbits that arise in the attractor of a dynamical system, which increases in complexity as a route to chaos. This is distinct from the definition of a QPO in astronomy identified by broad peaks in the power spectrum of a black hole light curve.} to chaos. A relationship between spectral state and dynamical variability characteristics may provide hints to the origins of the emission in each state and the mechanism responsible for state changes. 

Finally, we reiterate that the accretion state of a system has never been predicted using only the total-band flux variability information (i.e., without spectral information) and has yet to be directly connected to the temporal dynamics. Our goal is to demonstrate this method for the first time on a canonical black hole source like Cyg X-1 as a proof-of-concept that will then be employed for a larger sample in a subsequent study.

This paper is organized as follows: in Sec.~\ref{sec:data} we review the instrumentation used for collection and the type of data for this study. In Sec.~\ref{sec:methods}, we define the methods that are critical to performing the analysis of Cyg X-1, such as Hardness Intensity Diagrams, Recurrence Plots, and machine learning techniques including k-Nearest Neighbors and Random Forests. In Sec.~\ref{sec:analysis}, we present our results comparing spectral states to variability statistics. In Sec.~\ref{sec:conclusions}, we provide concluding remarks and highlight the necessary steps to continue this study, expanding our results to other sources and ultimately to large surveys to aid the discovery of the fundamental processes of spectral state changes.

\section{Data}\label{sec:data}

In this study, we consider the long-term behavior of Cyg X-1 as a function of time across a broad range of energies. The data utilized covers more than two decades with daily monitoring from two instruments: the decommissioned Rossi X-ray Timing Explorer (RXTE) All-sky Monitor (ASM) \citep{levine1996} and the ongoing Monitor for All-sky X-ray Image (MAXI) onboard the ISS \citep{matsuoka2009}. RXTE-ASM and MAXI each have 3 X-ray energy bands which we have labeled as soft, intermediate, and hard covering a total range from 1.5 – 20 keV.

The RXTE-ASM was in operation for approximately 16 years before it was decommissioned in early 2012. During its years of operation, it utilized 3 Scanning Shadow Cameras (SSCs) and a position-sensitive proportional counter (PSPC) to measure the intensities of different X-ray sources \citep{levine1996}. RXTE had an energy range of $2-200$ keV, but the ASM collected data in 3 bands: $1.5-3$ keV (`soft'), $3-5$ keV (`intermediate'), and $5-12.2$ keV (`hard'). This instrument would randomly select a source to scan between 5 and 10 times per day. We obtained the public RXTE-ASM data in 90-second dwells from the High Energy Astrophysics Science Archive Research Center (HEASARC).

The MAXI telescope was attached to the Japanese Experiment Module on the International Space Station (ISS) in 2009. It has two main X-ray slit cameras and two types of X-ray detectors. The Gas Slit Camera (GSC) has an energy range of $2-30$ keV and the other is an X-ray CCD, Solid-state Slit Camera (SSC) with a range of $0.5-12$ keV. One goal of this telescope is to measure the long-term variability of different X-ray sources \citep{matsuoka2009}. MAXI was made to be more sensitive, reaching $\sim$1 mCrab (at 1 week), than the RXTE ASM, which only had a sensitivity of about 10 mCrab. This allows MAXI to collect data from weaker X-ray sources but also continue monitoring some of the same sources as RXTE-ASM, such as Cyg X-1. We obtained the daily monitoring light curves by MAXI of Cyg X-1 from the public archive provided by the RIKEN, JAXA, and MAXI team\footnote{MAXI website: http://maxi.riken.jp/}.

\begin{figure*}
	\includegraphics[width=\textwidth]{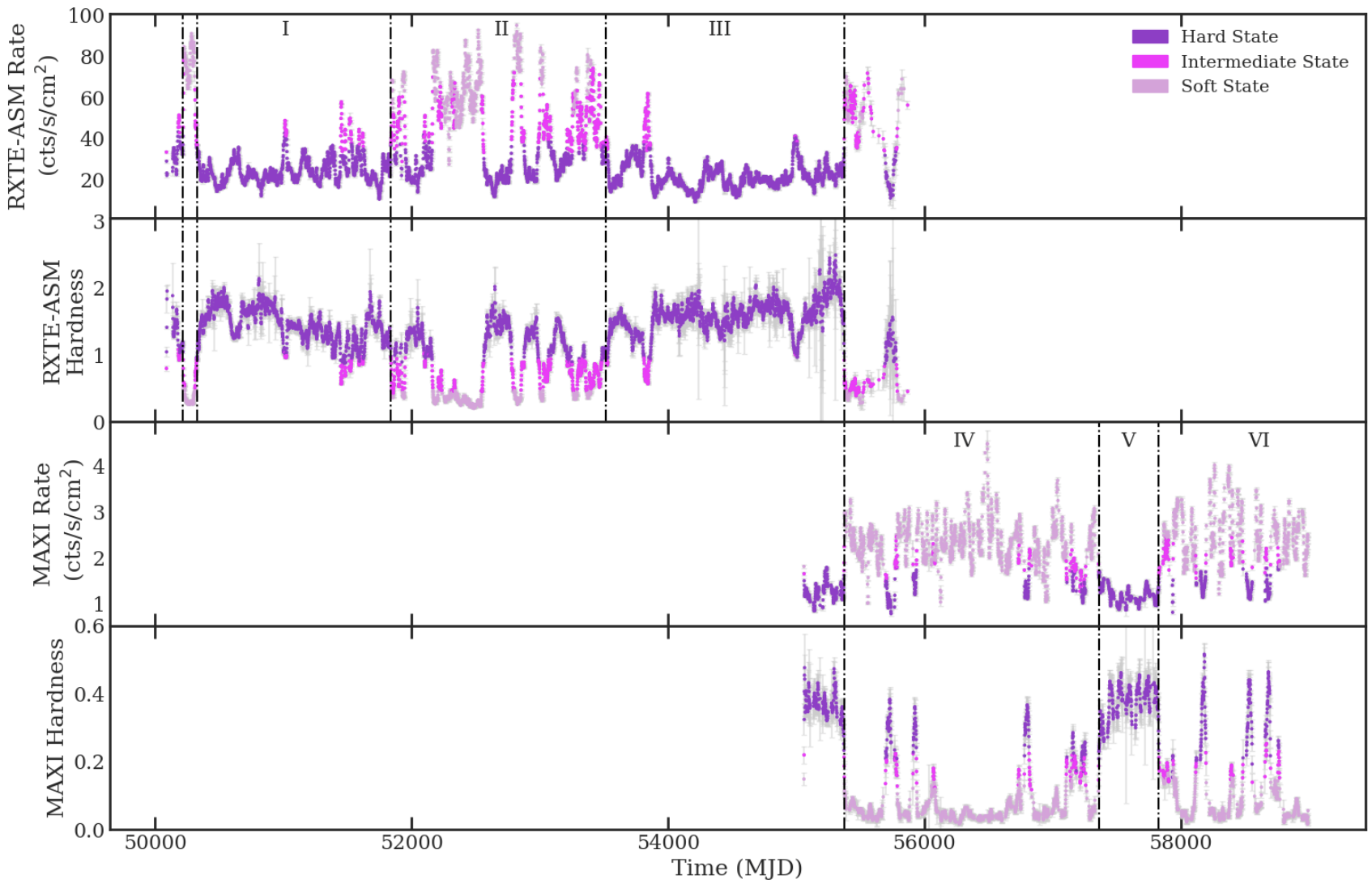}
    \caption{The RXTE-ASM and MAXI average daily monitoring light curves of Cyg X-1 (first and third panels, in energy bands $1.5-12.2$ keV and 2-20 keV respectively) and the hardness ratios over time (second and fourth panels). The hardness ratios are defined as ($5-12$ keV) / ($1.5-3$ keV) for RXTE-ASM and ($4-10$ keV) / ($2-4$ keV) for MAXI. The coloring of the light curves and hardness correspond to the classifications of soft (lavender/light), intermediate (pink), and hard (purple/dark) spectral states as defined in Sec.~\ref{sec:HIDs}. The vertical dot-dashed lines separate epochs of different behaviors as discussed in Sec.~\ref{subsec:behavior}. Errors for all features are represented in solid grey vertical lines and we note there are many times at which the errors are not large enough to be visible outside the marker.}
    \label{fig:lightcurves_separate}
\end{figure*}

The average daily monitoring of Cyg X-1 by RXTE-ASM and MAXI are individually displayed in Fig.~\ref{fig:lightcurves_separate} (top and third panels, respectively). The energy ranges of each bandpass for both RXTE-ASM and MAXI are presented in Table~\ref{tab:energy_ranges}.

In this study, we have analyzed data from RXTE-ASM and MAXI both separately and combined. In order to combine the data sets, we normalized both types of data to the Crab Nebula. During the overlapping times, we followed the method outlined in \cite{Phillipson2018} to scale the MAXI data to the ASM and subsequently convert the light curve into physical units of erg cm$^{-2}$ s$^{-1}$. This allowed us to view over 20 years of data collection together in the variable time series and subsequent analyses described in this paper.

\section{Methods}\label{sec:methods}
  
\subsection{Accretion State Classification: Hardness Intensity Diagrams}\label{sec:HIDs}

We use Hardness Intensity Diagrams (HIDs) for visualizing and generating the labels of the different spectral states of Cyg X-1 for use in the machine learning application. We define the RXTE-ASM hardness as the ratio of the $5-12$ keV to $1.5-3$ keV bands and the MAXI hardness as the ratio between the $4-10$ keV and $2-4$ keV bands. These definitions are based on the spectral classifications made of the daily ASM and MAXI monitoring by \cite{grinberg2013}. The hardness ratios as a function of time for RXTE-ASM and MAXI are presented in the second and fourth panels of Fig.~\ref{fig:lightcurves_separate}, respectively. 

\begin{figure}
	\includegraphics[width=\columnwidth]{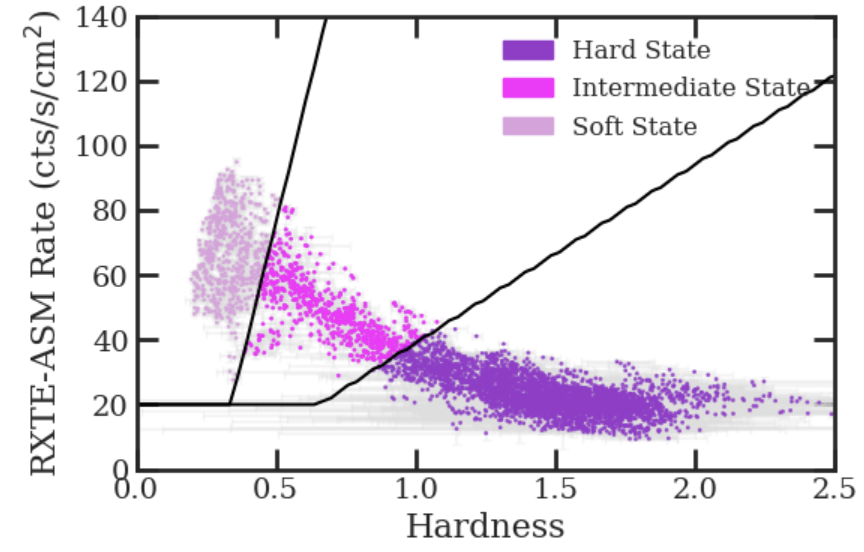}
	\includegraphics[width=\columnwidth]{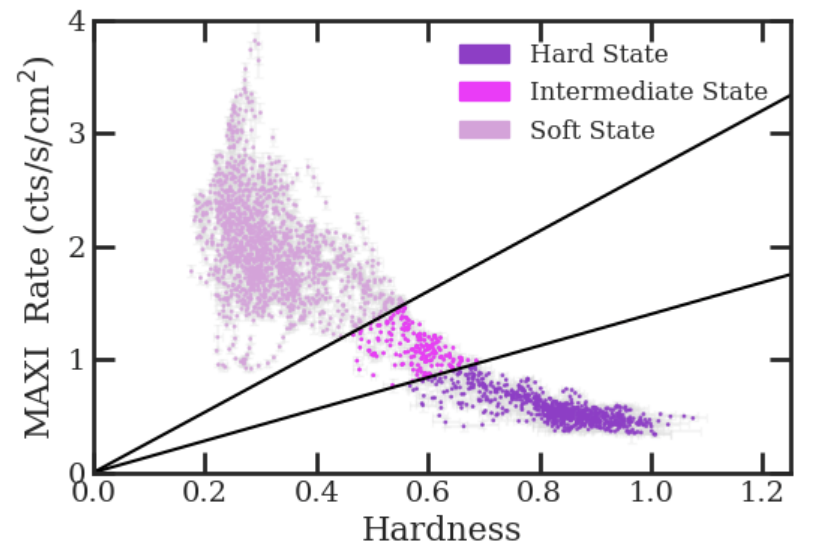}
    \caption{The hardness intensity diagrams Cyg X-1 from (top) the RXTE-ASM, with the hardness ratio defined as ($5-12.2$ keV)/($1.5-3$ keV) and (bottom) MAXI, with the hardness ratio defined as ($4-10$ keV)/($2-4$ keV). The solid black lines define the cuts defined by \protect\cite{grinberg2013}, from Table~\ref{tab:asm_maxi_cuts}. Each observation is classified as in either the soft, intermediate, or hard state and color-coded accordingly as light purple, hot pink, or dark purple, respectively.}
    \label{fig:HIDs}
\end{figure}

While hardness is typically inversely proportional to flux intensity, the correlation of spectral index to ASM count rate is complex and ASM or MAXI count rate alone is not reliable enough to separate between accretion states (\citealt{grinberg2013}, \citealt{Zdziarski2011}). Consequently, \cite{grinberg2013} implemented a scheme for direct classification of the ASM data using ASM observations that are simultaneous with spectra from pointed RXTE PCA observations. They identifed a relationship between full-band ASM count rate and individual energy bands with photon index. \cite{grinberg2013} employed a similar method of connecting spectra from pointed RXTE PCA observations with MAXI monitoring. The MAXI soft band count rate and the intermediate and soft energy bands are similarly correlated with spectral index. We use the resulting cuts made by \cite{grinberg2013} for both RXTE-ASM and MAXI to define the different accretion states of Cyg X-1 as a function of time. The mathematical expressions for these cuts can be found in Table~\ref{tab:asm_maxi_cuts} (a replication of Table 2 from \citealt{grinberg2013}). The resulting HIDs for RXTE-ASM and MAXI monitoring of Cyg X-1 are presented in Figure~\ref{fig:HIDs}, where the cuts distinguishing spectral states are visually represented by the solid black lines. 

\begin{table}
\caption {Bandpass Energy Ranges of Observations (in keV)} \label{tab:energy_ranges} 
    \begin{tabular}{lllll}
    \hline
    & Soft  & Intermediate  & Hard  & Full Band  \\
    \hline
    RXTE-ASM  & $1.5-3$ & $3-5$ & $5-12.2$ & $1.5-12.2$\\
    MAXI & $2-4$ & $4-10$ & $10-20$ & $2-20$ \\
    \hline
    \end{tabular}
\end{table}

\begin{table}
\caption{Spectral state definitions, as defined by Table 2 of \protect\cite{grinberg2013}. The RXTE-ASM full-band ($1.5-12.2$ keV) count rate or MAXI soft band (2-4 keV) count rate is represented by `$c$' and corresponding hardness of each instrument by `$h$', with a constant defined as $h_0 = 0.28$.} \label{tab:asm_maxi_cuts}
    \begin{tabular}{lllll}
    \hline
    & RXTE-ASM & MAXI \\
    \hline
    Hard   & $c \leq 20 \lor c \leq 55(-h_0) $ & $ c \leq 1.4h $ \\
    \\
    Intermediate & $c > 20 \land c \leq 55 (h-h_0)$  & $1.4h < c \leq 8/3h $ \\
    & $< c \leq 350 (h-h_0)$\\
    \\
    Soft & $c > 20 \land c > 350 (h-h_0)$ & $ 8/3 h < c $ \\
    \hline
    \end{tabular}
\end{table}

\subsection{Characterizing Timing Variability: Recurrence Plots}\label{methods:RP}

A Recurrence Plot (RP) is a graphical representation of a two-dimensional `recurrence matrix' containing all the nonlinear correlations in the light curve \citep{eckmann1987}. It is a tool that assists in the analysis of dynamical systems and is a method from nonlinear time series analysis \citep{marwan2007}. The recurrence matrix contains information on trajectories in the phase space of the system, a higher-order space that is typically constructed by a time series and its derivatives. For example, the phase space of a simple pendulum is a circle, tracing the angular position against angular velocity over time. 

The recurrence matrix correlates positions in time to the closeness of those positions in phase space. For a time series embedded in $m$-dimensional phase space ($\vec{x}_i \in \mathbb{R}^{m}$), the recurrence matrix is defined as:
\begin{equation}\label{eq:rp_eqn}
    \mathbf{R}_{i,j}= \Theta( \epsilon - \left| \right| \vec{x}_i-\vec{x}_j \left| \right| ), 
\end{equation}
where $\Theta$ is the Heaviside function, $\epsilon$ is a threshold distance to compare the `closeness' of the two states ($i$ and $j$) in phase space, $N$ is the number of observations in the time series, and $i,j=1,...,N$. For experimental or observational data, the threshold should be larger than five times the observational noise \citep{Thiel2002}, in order to avoid spurious recurrences, while not exceeding the maximum phase space diameter \citep{Zbilut1992}.

RPs are the visualization of the recurrence matrix where, for states that are close in phase space, Eq.~\ref{eq:rp_eqn} returns unity, and zero otherwise. RPs are therefore binary images and symmetric about the main diagonal, called the line of identity. Examples of the RPs of various canonical systems are presented and discussed in Appendix~\ref{sec:app_rps}. RPs can be constructed for univariate time series, such as light curves, by reconstructing phase space from an appropriate embedding. For known dynamical systems, the direct time derivatives of the scalar observable can be used (for example, the observable and its first derivative for a 2-dimensional phase space). For unknown dynamical systems, there are several methods for phase space reconstruction. The time delay method is the most common and is used for studying the full RP of Cyg X-1. Discrete Legendre polynomials can be used to reconstruct the derivatives of the system and are used for constructing the RPs in the machine learning models developed in Sec.~\ref{sec:analysis}. We summarize each method of phase space reconstruction in the Appendix~\ref{sec:phase_space}.

\begin{figure}
	\includegraphics[width=0.5\textwidth]{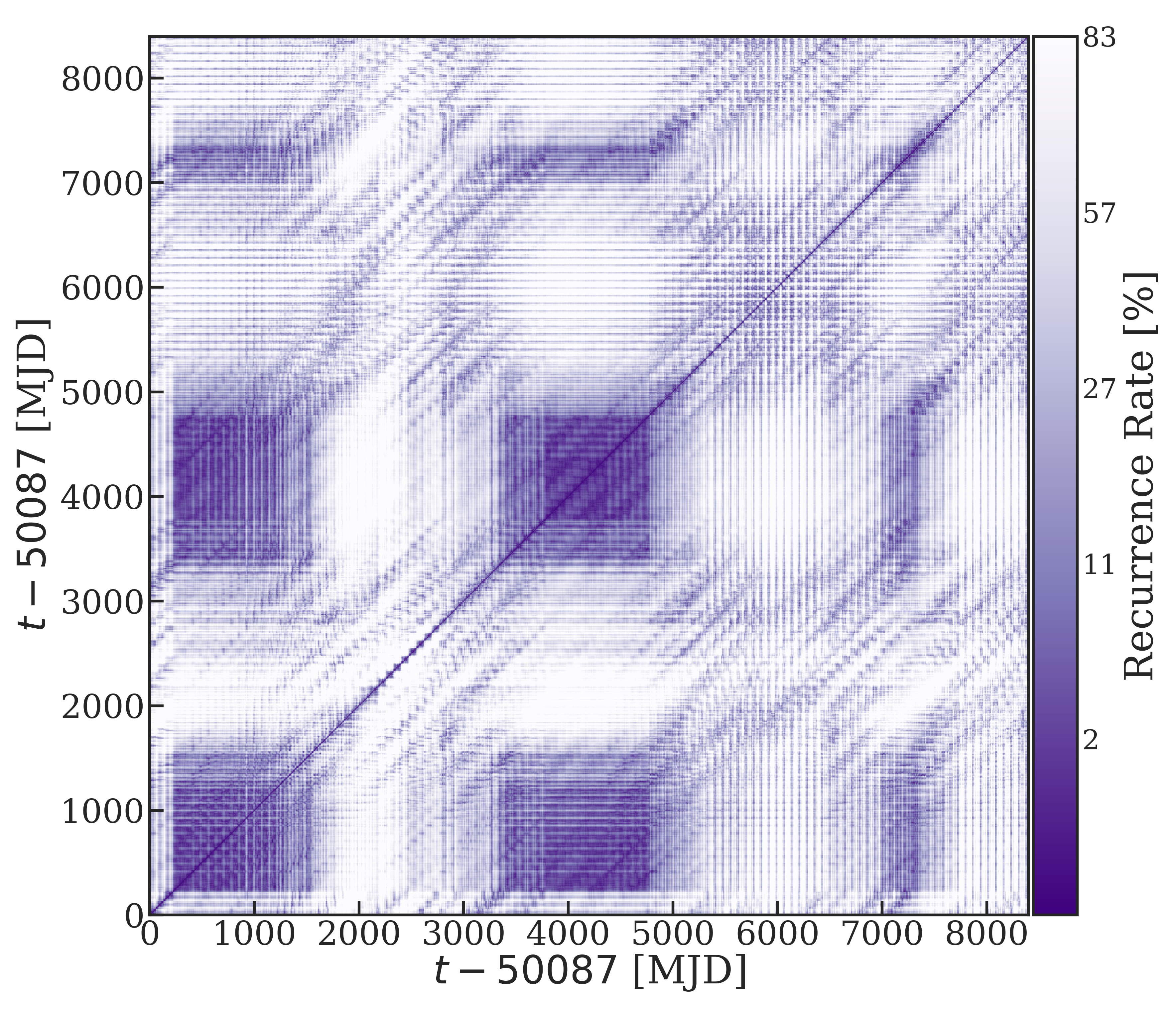}
    \caption{The un-thresholded RP of Cyg X-1. The recurrence rate that corresponds to the threshold, $\epsilon$, from Eq.~\ref{eq:rp_eqn} is replaced with a purple color bar to represent the recurrence points. This version of the RP is sometimes referred to as the `distance matrix,' as it represents the relative distances of every pair of points in phase space. Darker colors correspond to smaller thresholds, indicating the phase space positions for two points in time are near to each other. Lighter colors correspond to larger thresholds, indicating these positions in phase space are far from each other.}
    \label{fig:full_RP}
\end{figure}

If we replace the threshold, $\epsilon$, in Eq.~\ref{eq:rp_eqn} with a colorbar, we obtain an `un-thresholded' RP, displayed in Fig.~\ref{fig:full_RP} for Cyg X-1. The un-thresholded RP of Cyg X-1 contains multiple qualitative features. The diagonally-oriented striations represent times when the light curve is close to repeating itself. The overall block-like plaid features of the RP indicate times when the light curve is undergoing significant variations over the long term. In general, very faded regions would indicate a source is non-stationary. Other qualitative features unique to various dynamical systems can be seen in the examples given in the Appendix (Fig.~\ref{fig:rps_examples}). 

\subsubsection{The Time-Dependent RP}\label{sec:time_dependent_rps}

The RP can be constructed for the entire length of a time series or in a sliding window in which one obtains an RP as a function of time, called a \textit{windowed recurrence plot}. A windowed RP enables one to determine recurrence features as a function of time and can be used for determining dynamical regime changes. Specific RP features are sensitive to changes in the underlying dynamical system that generates the time series when computed in a windowed RP. In particular, the average vertical line length (called the `trapping time,' $TT$) is sensitive to dynamical transitions between periodic windows and chaotic regimes (and is zero for purely periodic dynamics; \citealt{marwan2007}). Vertical lines also occur much more frequently in regions of intermittency than in other chaotic regimes and thus increase substantially at these points. Secondly, the ratio of determinism to the recurrence rate ($DET/RR$) can detect state transitions in which the recurrences re-organize themselves to and from diagonal line structures (or to/from a more ordered state; \citealt{Webber1994}). An increase or decrease in $DET/RR$ thus locates times at which the dynamics of the underlying system fundamentally shift. The definitions of these RP features and the others used in the machine learning methods are provided in Appendix~\ref{sec:rqa}.

\subsection{Machine Learning Models}

Two machine learning (ML) techniques are used to find a correlation between the RP features (definitions detailed in Appendix~\ref{sec:rqa}) as a function of time in the combined light curve of Cyg X-1 and the spectral states defined by the HIDs in Fig. ~\ref{fig:HIDs}. We employ supervised learning using both training (labeled predictor variables) and testing (unlabeled target variables) data sets. The training data is used to tune the model to create the best and most accurate results possible for predicting the testing data. Classification algorithms predict discrete class labels, whereas regression predicts continuous values. We use the supervised learning methods of Random Forests and k-Nearest Neighbors regression algorithms in this study. 

\subsubsection{k-Nearest Neighbors}
 
k-Nearest Neighbors (KNN) is a memory-based prototype method that uses proximity to other points to predict a classification label \citep{hastie2017}. It is one of the most popular supervised learning methods for classification based on its simplicity. With an initial query point, $x_0$, KNN determines the $k$ closest points in the training data set to $x_0$ based on Euclidean distance. Once the `k' neighbors are found ($x_{r}$, for $r=1,...,k$), the regression algorithm uses the mean of the neighbors ($x_{r}$) to determine the class label of each point. The mean prediction can also be weighted by the distance of each neighbor to the initial query point. 

\subsubsection{Random Forests}

A Random Forest (RF) is considered an ensemble method, which produces one predictive model utilizing multiple base estimators \citep{Pedregosa2011}, such as an ensemble of Decision Trees \citep{Quinlan1986}. Decision trees perform classification by asking simple questions to separate the data via a binary split. This begins with one rule based on the numerical values of the model's features and preferentially divides the data set into two similarly sized sections --- one that satisfies the rule and one that does not. Each split point is called a node and each categorization is a branch of the decision tree. 

At each node of a decision tree, the algorithm will select a set of variables, such as the recurrence features from an RP, and then choose the best-split point by using an impurity or loss function \citep{Pedregosa2011}. For our case of the decision tree regressors, where the predicted class is a continuous variable, we choose parameters and node splits that minimize the mean squared error of the ending classification.

RFs are a form of probabilistic classification that decrease variance and reduce the risk of over-fitting by averaging multiple decision trees \citep{Breiman2001}, called a forest. An RF will randomly select a subset of the training dataset and a random selection of the features (predictor variables) to construct each decision tree. Specifically, the first step in an RF algorithm is to draw bootstrap samples from the training data to create multiple decision trees, $T_b$, for a forest size of $b=1$ to $B$ for each sample. Together, these trees create a forest, $\{T_b\}^B_1$. It is then possible to predict values with either regression (using the mean) or classification (via majority vote). Combining the predictions of randomly constructed decision trees into one best estimator using an RF prevents overfitting and increases accuracy over a single tree \citep{vanderplas}.

\section{Analysis and Results}\label{sec:analysis}

\subsection{General Behavior}\label{subsec:behavior}

We embed the full-band RXTE-MAXI light curve of Cyg X-1 into phase space using the time delay method (detailed in Appendix~\ref{sec:phase_space}) and generate a windowed RP with width 2000 days in 1-day increments across the entire light curve. We use a threshold corresponding to a 5 per cent recurrence rate for the full RP, which results in recurrence rates ranging between 10 per cent and 80 per cent for the individual sub-RPs that constitute the entire windowed RP. 

\begin{figure*}
	\includegraphics[width=\textwidth]{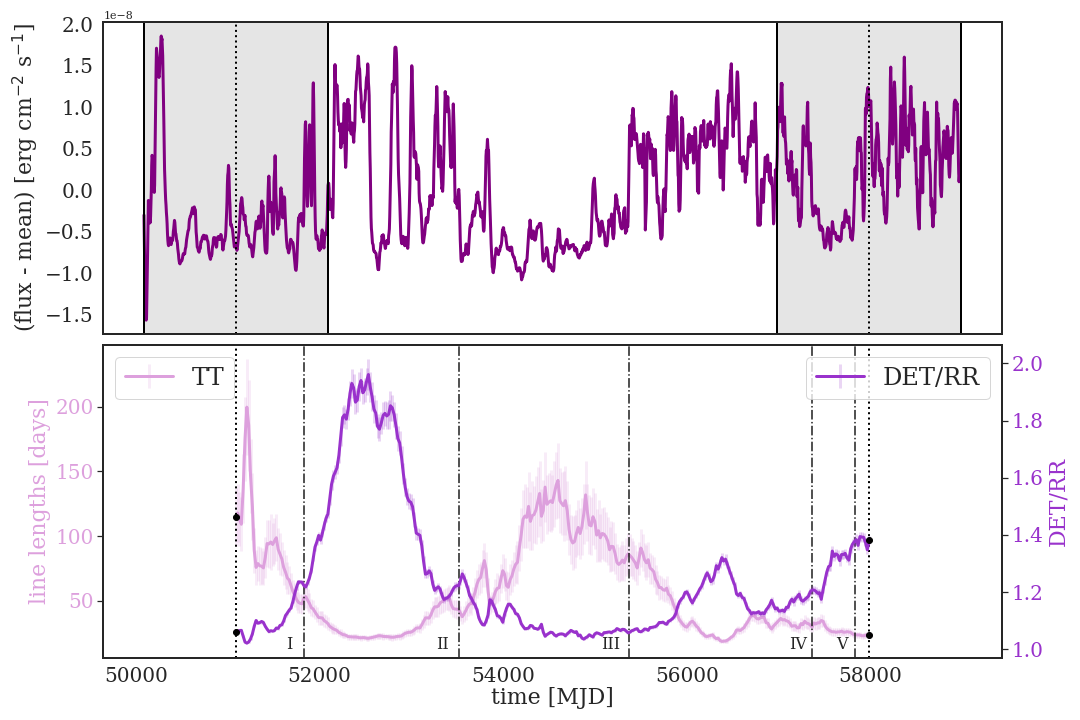}
    \caption{Top panel: The combined, mean-subtracted full-band flux (covering 2-20 keV) from RXTE-ASM and MAXI (errors are omitted from this figure for clarity, but in general do not exceed the width of the line used for the light curve). Bottom panel: Two features derived from the windowed RP that measure laminarity (the trapping time, $TT$, in light purple) and determinism ($DET/RR$ in dark purple) over time. The gray shaded windows and vertical dotted line in the top panel show the width and center time, respectively, of the windowed RP at the start and end of the light curve, with the center times marked by vertical dotted lines in the bottom panel for reference. The dot-dashed vertical lines separate the six epochs of different behavior seen in the hardness of Fig.~\ref{fig:lightcurves_separate} and described in comparison to the RP features in Sec.~\ref{subsec:behavior}. In the bottom panel, the 95 per cent confidence bounds are presented as vertical lines in lighter colors of the corresponding $TT$ or $DET/RR$ features.}
    \label{fig:lightcurve_DET_vs_TT}
\end{figure*}

From the windowed RP, we compute the average vertical line length, $TT$, and the ratio of determinism to recurrence point density, $DET/RR$, both of which are features sensitive to dynamical regime changes. Since we construct the windowed RP in 1-day increments, we obtain $TT(t)$ and $DET/RR(t)$ with daily cadence. The combined RXTE-ASM and MAXI light curve of Cyg X-1 and the corresponding changes in $DET/RR$ and $TT$ over time are displayed in Fig.~\ref{fig:lightcurve_DET_vs_TT}. We establish 95 per cent confidence bounds on both RP features using the structure-preserving bootstrap resampling technique developed by \citealt{Schinkel2009} for use with RPs. We adapt the code developed by \citealt{Schinkel2009} as part of the \textit{MATLAB}-based \textit{CRP Toolbox} (RP software written by \citealt{Marwan2002} and \citealt{marwan2007}) into Python to generate the windowed RP, the $DET/RR$ and $TT$ features, and their confidence bounds. 

We define six epochs of time (labeled I-VI in Fig.~\ref{fig:lightcurves_separate} and Fig.~\ref{fig:lightcurve_DET_vs_TT} with date ranges defined in Table \ref{tab:epoch_MJD}) that demonstrate distinct spectral behavior, in order to compare the change in the $TT$ and $DET/RR$ features with spectral states. To define these epochs, we determined the points at which Cyg X-1 enters or leaves an extended period of continuous behavior. For example, as can be seen in Fig.~\ref{fig:lightcurves_separate}, Cyg X-1 exits the soft state at MJD 50326 and does not return until MJD 51837. 
 
 \begin{table}
      \caption{Epochs of similar behavior, visualized in Fig.\ref{fig:lightcurves_separate}. }
     \centering
     \begin{tabular}{c|c}
     Epoch & Approximate MJD range \\
     \hline
     I & 50326-51837\\
     II & 51838-53519\\
     III & 53520-55375\\
     IV & 55376-57362\\
     V & 57363-57831\\
     VI & 57832-58995\\
     \end{tabular}
     \label{tab:epoch_MJD}
 \end{table}

Epochs I, III, and V have a high occurrence of the hard spectral state, while in Epoch II Cyg X-1 undergoes frequent state changes between the HSS and LHS, also observed in Epochs IV and VI, albeit to a lesser extent. Between MJD 50350–51000, \cite{grinberg2013} found 99 per cent of RXTE-ASM observations to be in the hard state, a time period that most closely aligns with the first half of epoch I. From MJD 51000 to MJD 53900, \cite{grinberg2013} found 63 per cent of observations to be in the hard state, which subsequently increases to 97 per cent between MJD 53900 and MJD 55375 (comparable to epochs II and III, respectively). Finally, it was found that 75 per cent of MAXI observations were in the soft state, similar to epoch IV, between MJD 55375 and MJD 56240. The hard and soft states are very stable compared to the intermediate state, with the hard state being the most stable of the three. Per \cite{grinberg2013}, Cyg X-1 stays in the hard state for at least one week over 85 per cent of the time and is steadily in the soft state for one week 75 per cent of the time. In contrast, the source exists in the intermediate state only transiently, with a 50 per cent chance of Cyg X-1 remaining in the intermediate state within a 3-day period. 

A significant increase or decrease in $DET/RR$ can be used as a probe of changes in the variability pattern, as explained in Sec.~\ref{sec:time_dependent_rps}. Indeed, we see that in epochs II, IV, and VI, the hardness of Cyg X-1 varies between soft and hard more frequently, with an increased amount of time spent in a soft state. In contrast, in epochs I, III, and V, Cyg X-1 is predominantly evolving in the hard state. During these epochs, we observe a dramatic increase in the trapping time, $TT$, which is a measure of the average length of a vertical line segment in the RP. $TT$ is sensitive to dynamical state transitions between periodic windows and chaotic regimes, in which low values correspond to periodic dynamics and higher values correspond to a regime of chaotic intermittency. Intermittency occurs when a signal experiences irregular bursts of chaotic behavior amidst otherwise laminar phases. Our initial interpretation suggests that $DET/RR$ detects changes between Cyg X-1 existing in the LHS and times when it undergoes outbursting behavior and frequent spectral state changes. Similarly, $TT$ appears to be highly correlated with the hardness of Cyg X-1 and suggests that the accretion properties in the LHS lead to intermittent or chaotic dynamics in the inter-day variability.

\subsection{Correlations between RP Features and Spectral State}\label{subsec:correlations}

The qualitative differences between the six epochs demonstrate that the RP features may evolve in time and be correlated to the spectral state. To quantify this behavior, we embed the full-band $ASM$-$MAXI$ light curve of Cyg X-1 into 3-dimensional phase space using the Legendre coordinates method and generate a windowed RP with five different window widths (2000, 1000, 500, 250, and 100 days) in 1-day increments across the entire light curve. We subsequently computed eight of the RP features for each sub-RP in the windowed RP: $DET$, $LAM$, $ENTR$, $L_{mean}$, $TT$, $L_{max}$, $V_{max}$, and $DIV$ (complete definitions can be found in Appendix~\ref{sec:rqa}).

We compare the mean values of each RP feature in the soft and hard spectral states for every RP window size explored in order to determine which RP features are most correlated to the spectral state. That is, for each window size, we computed the average $DET$ for each sub-RP weighted by the fraction of time spent in either the hard or soft state to distinguish the mean $DET$ for each state. Time spent in the intermediate state is significantly less than in either the soft or hard states, as discussed in Sec.~\ref{subsec:behavior}, and exists only transiently. We therefore only consider the hard and soft states for comparison. To determine the significance of a difference in means between spectral states, we use Welch's $t$-test \citep{Welch1947}, a modification of the standard Student's $t$-test \citep{Ttest1908} that accounts for unequal variances or sample sizes. 

The sample sizes in each spectral state are highly skewed, with over half the observations in the hard state and the remainder split between the soft and intermediate states. The sub-RPs also contain overlap with each other, since the windowed-RP involved incremental steps of one day for each sub-RP. We therefore chose to randomly sub-sample the recurrence features from 150 sub-RPs at a time in each spectral state, perform the $t$-test and compute the corresponding p-value, for 1000 sets of sub-samples. This process enabled us to satisfy the assumptions of Welch's $t$-test, that the populations being tested represent random and independent samples from a normal distribution, and to mitigate potential error due to uneven sample sizes.

The resulting p-values from the $t$-test comparing the mean RP features between the hard and soft spectral states are summarized in Table~\ref{tab:t-test_fractional}. We also include tables comparing the intermediate state to the soft and hard states in Appendix~\ref{sec:pvalues}, though no significant differences were found. This suggests that individual RP features are not sufficient to distinguish the intermediate state from the other spectral states and that a more sophisticated ML model may be required.

\begin{table}
\caption {The p-values obtained from Welch's $t$-Test comparing the means of RP features between the hard and soft spectral states. The sub-RPs used to compute the RP features were classified as in either the hard or soft spectral state by using the weighted mean of the RP features, where the weights were defined as the fractional amount of time spent in either the hard or soft state. For those p-values that we consider significant ($p \le 0.05$), we apply a subscript: an `H' indicates that the average feature value was greater in the hard state and a `S' indicates the higher average value is in the soft state.} \label{tab:title}
\begin{tabular}{ |c|c|ccccc|} 
\hline
 & \multicolumn{5}{|c|}{Window Size (days)} \\
\hline
Feature       &  2000 & 1000 & 500  & 250  & 100\\
\hline
Determinism:  & 0.07  & 0.07 & 0.05$^H$ & 0.04$^H$ & 0.56 \\
Laminarity:   & 0.22  & 0.11 & 0.02$^H$ & 0.12 & 0.55\\
Entropy:      & <0.01$^H$ & 0.02$^H$ & 0.03$^H$ & 0.10 & 0.49\\
L Mean:       & 0.03$^H$  & 0.06 & 0.07 & 0.07 & 0.47\\
TT:           & 0.02$^H$  & 0.04$^H$ & 0.02$^H$ & 0.17 & 0.53\\
L Max:        & 0.15  & 0.12 & 0.19 & 0.32 & 0.43\\
V Max:        & 0.56  & 0.20 & 0.06 & 0.38 & 0.35\\
Divergence:   & 0.07  & 0.02$^S$ & 0.13 & 0.14 & 0.47\\
\hline

\hline
\end{tabular}\label{tab:t-test_fractional}
\end{table}

In general, we find that most RP features are systematically higher in the hard state than in the soft state, with a notable exception being the $DIV$ feature (the inverse of the longest diagonal line length). We also found that the Shannon Entropy ($Entropy$) and Trapping Time ($TT$) were the most common significant features across a range of RP window sizes. There was no significant difference between the RP features once the window size was 100 days, but some were still marginally significant (e.g., p-values less than 0.1) in the 250-day window (or at least close to the threshold). This suggests that the 250-day window is the smallest window possible that will still have distinguishable features between spectral states over the long term.

The determinism and lamanarity also retain relatively low p-values, though they are only highly significant for certain window sizes. The p-values for the lamanarity decrease with decreasing window size, before increasing substantially in the 100-day window. The determinism p-values display a similar effect, though to a lesser extent; indeed, the p-values remain marginally significant for all window sizes except the 100-day window. The p-values of all other recurrence features appear to steadily worsen with decreasing window size. This suggests that the lamanarity, in particular, is sensitive to window size and may be a feature that is not invariant for distinguishing unique features between RPs. 

To further clarify our results, we performed the non-parametric Mann Whitney U test (also known as the Wilcoxon Rank Sum test; \citealt{MannWhitney}), which tests the null hypothesis that the distributions underlying two samples are the same and makes no assumptions about the two samples being drawn from normal distributions. For the Mann Whitney U test we also found that the entropy, mean diagonal, and mean vertical line lengths resulted in significant p-values for a range of window sizes, rejecting the null hypothesis and supporting the alternative that the RP features are drawn from significantly different distributions. 

The significant p-values signify that the recurrence features are, in general, all higher in the LHS. 

We deduce that the RP contains recurrences in the phase space trajectory of the light curve that last longer, on average, and evident more deterministic behavior in the LHS. This could be interpreted as an LHS light curve containing more memory, or as fluctuations that persist for longer in an impulse-response system. However, the $TT$ feature, in particular, is a probe of dynamical transitions in addition to tracing laminar behavior: higher values indicate regions of chaotic intermittency (as opposed to periodic dynamics). An increase in the average diagonal line length is a more direct probe of the memory in the light curve, which traces how long a section of the light curve mirrors another section in the light curve when projected in phase space. Both longer vertical and diagonal line lengths appear in the LHS. The Shannon entropy measures the information, or uncertainty, associated with the physical process described by the probability distribution of the light curve and is computed from the distribution of diagonal line lengths in the RP. Higher values indicate light curves with high information entropy and uncertainty in the physical process, which may also indicate chaos. 

Thus, we posit that when Cyg X-1 is evolving in the LHS, it is more likely to exhibit unpredictable and potentially chaotic behavior than in the HSS. Furthermore, these dynamical differences are significant enough that the RP can be used as a probe for distinguishing between the soft and hard spectral states.


\subsection{Constructing Machine Learning Models from RP Features}\label{subsec:classifying via RP}

Our initial qualitative observations from Sec.~\ref{subsec:behavior} and the quantitative differences in RP features between states suggest that the RP features can be used to predict the spectral states of Cyg X-1. We, therefore, explore two popular ML methodologies (RFs and KNN) for predicting the spectral state of Cyg X-1 based on the RP features alone. One goal of this analysis is to determine whether the RP is significantly different in the three spectral states defined by \cite{grinberg2013}. Distinct differences in the RP between accretion states would suggest that the light curve contains distinct dynamics in each state.

A secondary goal of the analysis is to determine the minimum size RP required to distinguish differences between the spectral states over long-term monitoring. Given that Cyg X-1 will evolve in the HSS or LHS for many days at a time but also has a history of frequent state transitions on the order of days to months, we aim to acquire a functional ML model with high accuracy for small RPs that maximize time spent in a particular accretion state. Our results from Sec.~\ref{subsec:correlations} suggest that a window size of 100 days may be too small to distinguish features between the soft and hard spectral states.

We follow a similar procedure to Sec.~\ref{subsec:correlations} to develop the training sets used in the ML models and embed the full-band $ASM$-$MAXI$ light curve of Cyg X-1 into 3-dimensional phase space using the Legendre coordinates method. We generate a windowed RP with the five different window widths in 1-day increments across the entire light curve using a fixed recurrence rate of 10 per cent and compute the eight RP features for each sub-RP.

Regression models require the predicted class labels to be continuous values. We use the fractional amount of time that is spent in each spectral state in a given sub-RP as our class labels. During the times that the $ASM$ and $MAXI$ monitoring overlap, we use the spectral classification from $MAXI$.

Our full dataset consists of the eight recurrence features as a function of time and the fractional amount of time spent in each spectral state. Our expectation is that the regression models should characterize the overall texture differences between sub-RPs in different spectral states, where individual sub-RPs could be taken out of the context of the full light curve and regarded as a representation of an RP of that particular dominant spectral state. We would therefore consider different features of the sub-RPs to be more directly representative of particular spectral states, rather than merely as detections of state transitions (as indicated by $TT$ and $DET/RR$ in Fig.~\ref{fig:lightcurve_DET_vs_TT}).

To create the training and testing datasets, we perform a 70-30 split. That is, the training set is comprised of 70 per cent of the full dataset of recurrence features and spectral classifications, with the remaining 30 per cent set aside to assess the accuracy of the models. The sample sizes that constitute each class are uneven. Over half of the observations of Cyg X-1 are when it is in the LHS (52 per cent of the time), followed by about 30 per cent in the HSS, and the remainder in the intermediate state. We ensure that both our training and testing datasets retain the same proportions in each spectral state and do not significantly overlap. The recurrence features that were used as the predictor variables were normalized using the Z-score (zero mean and standard deviation of one).

Both the KNN and RF algorithms contain free parameters that the user selects before the learning process is implemented, called hyperparameters. To determine the optimal hyperparameters, we employ a grid-based cross-validation technique that systematically explores every combination of parameters within specified ranges. 

For the RF models, 5 parameters were optimized. The depth of each decision tree was limited to a choice between 4, 6, 8, 16, 50, 100, and 200, or set to no limit to the number of levels within each tree. The number of rules for each split/node ranged between 2 and 30, and the number of features to consider for each split was between 2 and 8. The use of a bootstrapping method to sub-select data for each decision tree was either applied or not applied, and the option for the number of trees in the forest included 25, 50, 100, 250, 500, and 1000 different trees. 

For the KNN models, we explored 2 parameters. The number of neighboring points to consider in the classification ranged between 2 and 50, and the weight of each neighboring point was either uniform or based on the Euclidean distance to the neighboring point. 

To select the optimal hyperparameters, we chose the model with the highest accuracy score (as implemented in the \textit{SciKit-Learn} package of Python). The accuracy score was based on the coefficient of determination, $R^2 = (1 - u/v)$, where $u$ is the residual sum of squares between the predicted and true labels of the data, and $v$ is the total sum of squares of the true labels of the data. While the tuning of hyperparameters is important for optimizing the accuracy of the final model, in general, all models were relatively robust against changes to the parameters. The range of accuracy scores for the hyperparameters of each model was within 10 per cent of each other.


\subsection{Model Assessment and Accuracy}
\label{subsec:assessment and accuacy}

Once we determine the optimal hyperparameters for each of our 20 models of varying window widths and ML algorithm choice, we evaluate two metrics for determining the accuracy of each model for comparison to each other. First, we compute the root-mean-square-error, or $RMSE$. For normalized predicted labels (e.g., those that exist on the unit scale), the $RMSE$ can be directly mapped to a percent error and are presented in Table~\ref{tab:regressor_scores}.

For all models, we also calculate the AUC-ROC score (which is the abbreviation for the `area under the curve' of the `receiver operating characteristic' graph), a common metric for determining the viability of models in data science. The ROC depicts the relative trade-offs between the true positive and false positive rates of a classifier. The AUC-ROC score varies between 0 and 1, with 1 being a perfect classifier (no false positives and all correct classifications are found). The AUC-ROC score is typically only computed for classification algorithms. Appendix~\ref{sec:AUC_ROC_Score} details the computation of the AUC-ROC score for our regression models.

\begin{table*}
\caption {Accuracy scores for the RF and KNN Regression models. The predictor variables are generated from the RP features of the time-dependent RP and the class labels are generated using the fractional amount of time spent in each spectral state in each sub-RP. The AUC-ROC score comparing each spectral state to the other two states, defined in Sec.~\ref{subsec:assessment and accuacy}, and the root-mean-square-error (RMSE), describing the percent error of the classified sub-RPs, are presented.} \label{tab:regressor_scores}
\begin{tabular}{ |c|c|ccccc|} 
\hline
& & \multicolumn{5}{|c|}{Window Size (days)} \\
\hline
Model & & 2000& 1000 & 500 & 250 & 100\\
\hline
\multirow{7}{2em}{RF}
&\multirow{3}{8em}{AUC-ROC Score} & Soft: 0.99 & Soft: 0.98 & Soft: 0.96 & Soft: 0.91 & Soft: 0.80 \\
& & Int: 0.99 & Int: 0.98 & Int: 0.96 & Int: 0.91 & Int: 0.79 \\
& & Hard: 0.995 & Hard: 0.99 & Hard: 0.96 & Hard: 0.91 & Hard: 0.80 \\
\\
&\multirow{3}{8em}{RMSE (\%)} & Soft: 0.5 & Soft: 1.0 & Soft: 3.0 & Soft: 7.0 & Soft: 14.3\\
& & Int: 0.4 & Int: 0.9 & Int: 2.5 & Int: 5.3 & Int: 11.7\\
& & Hard: 0.4 & Hard: 1.0 & Hard: 3.2 & Hard: 7.5 & Hard: 14.6 \\
\hline
\multirow{7}{2em}{KNN}
&\multirow{3}{8em}{AUC-ROC Score} & Soft: 0.996 & Soft: 0.998 & Soft: 0.99 & Soft: 0.95 & Soft: 0.82 \\
& & Int: 0.995 & Int: 0.997 & Int: 0.99 & Int: 0.95 & Int: 0.80 \\
& & Hard: 0.99 & Hard: 0.998 & Hard: 0.99 & Hard: 0.95 & Hard: 0.82 \\
\\
&\multirow{3}{8em}{RMSE (\%)} & Soft: 0.6 & Soft: 0.5 & Soft: 2.5 & Soft: 6.1 & Soft: 14.3\\
& & Int: 0.3 & Int: 0.2 & Int: 1.5 & Int: 4.7 & Int: 12.4\\
& & Hard: 0.5 & Hard: 0.5 & Hard: 2.6 & Hard: 6.5 & Hard: 14.6\\
\hline
\end{tabular}
\end{table*}

\begin{figure}
    \includegraphics[width=\columnwidth]{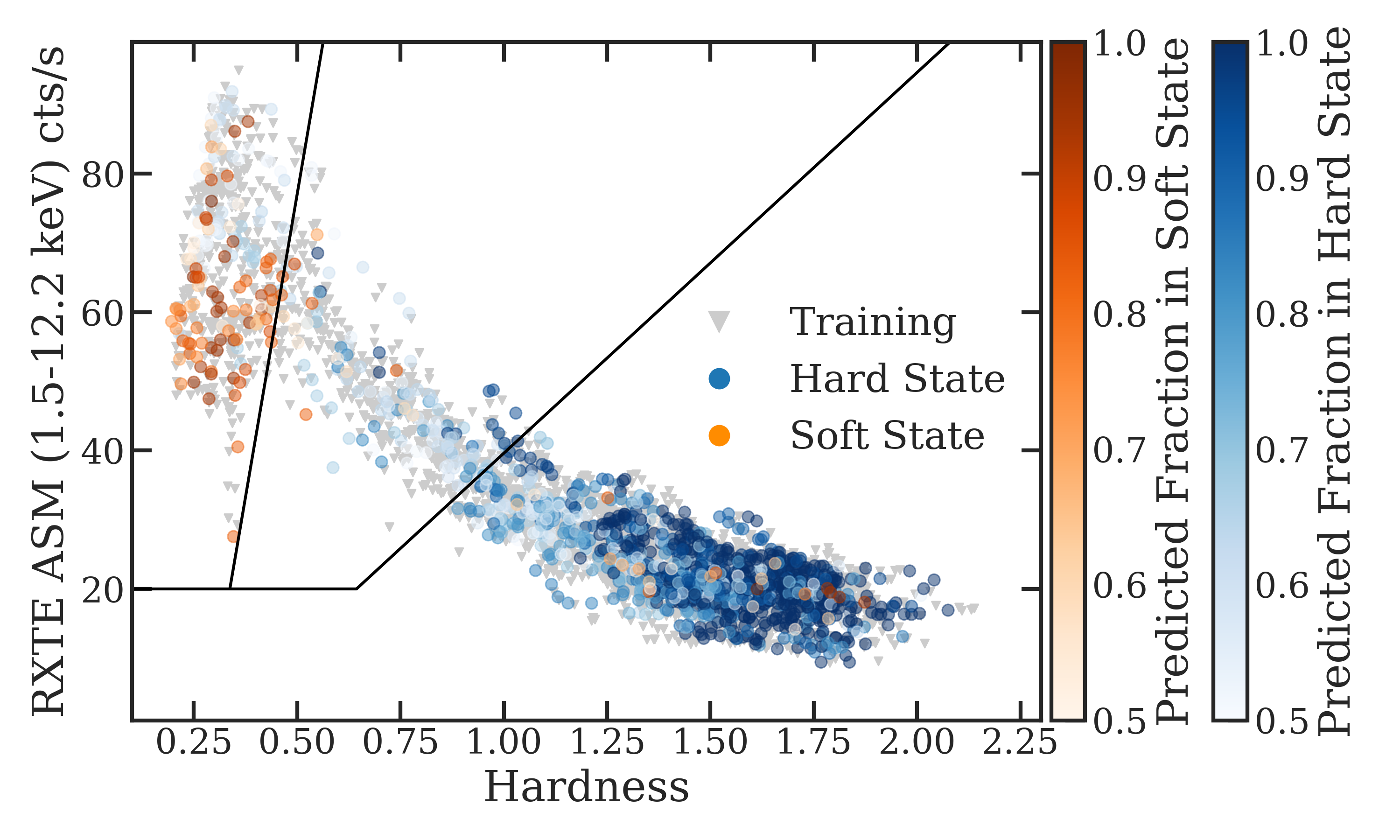}
	\includegraphics[width=\columnwidth]{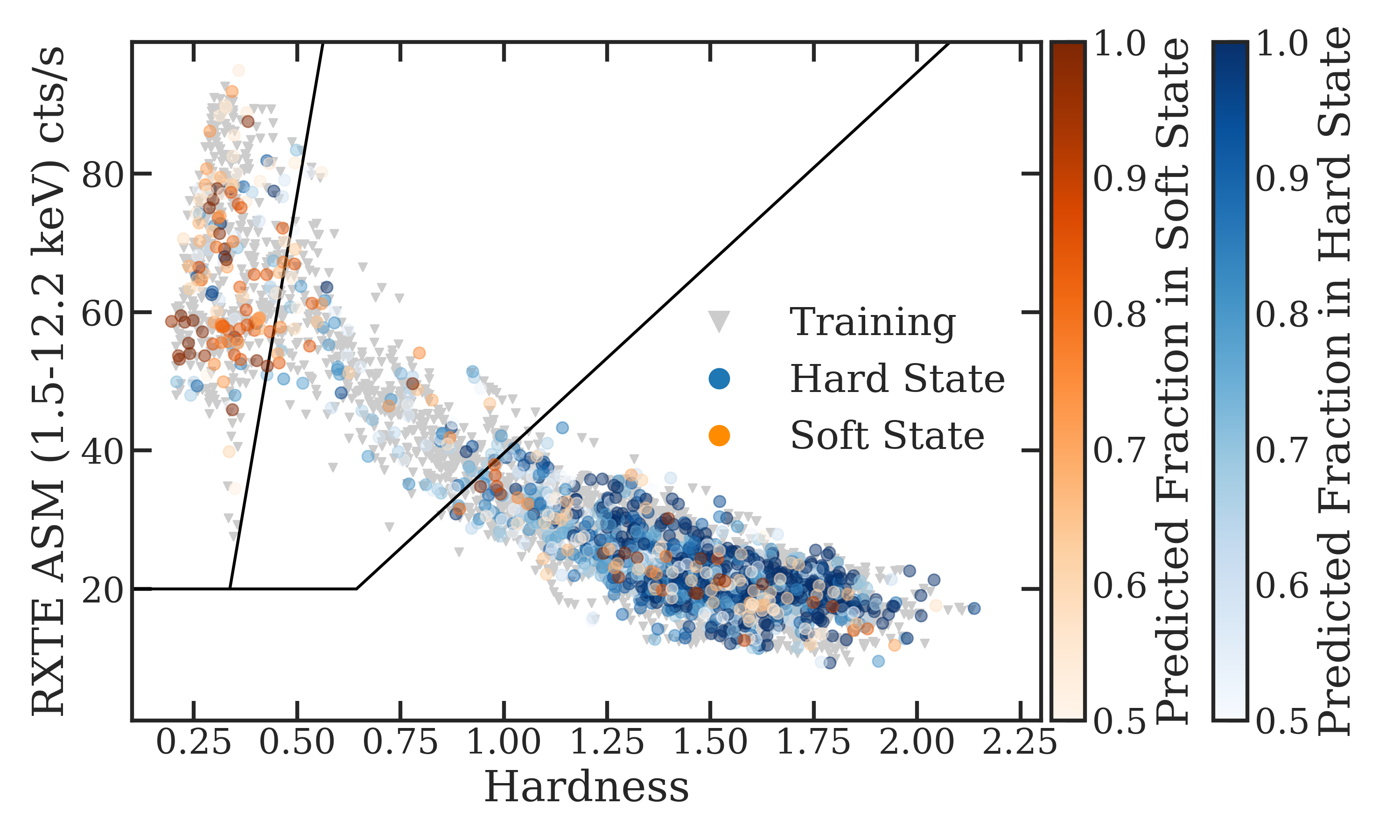}
    \caption{The RXTE-ASM HIDs, similar to Fig.~\ref{fig:HIDs}, colorized with the predicted values from the KNN regressor with a 250 d (top) or 100 d (bottom) RP window. The KNN regressor produces a prediction of the fraction of the RP that is in either the hard, intermediate, or soft states. For clarity, only the times for which the RP is at least 50 per cent in one of the states are colorized, with circle markers, and the intermediate state is not shown. The training data is represented by light gray triangles. Note that the darker the blue/orange, the more certain the model is that an observation is in the hard/soft state, corresponding to a greater fraction of time spent in either state.}
    \label{fig:HID_RXTE_predicted}
\end{figure}

\begin{figure}
    \includegraphics[width=\columnwidth]{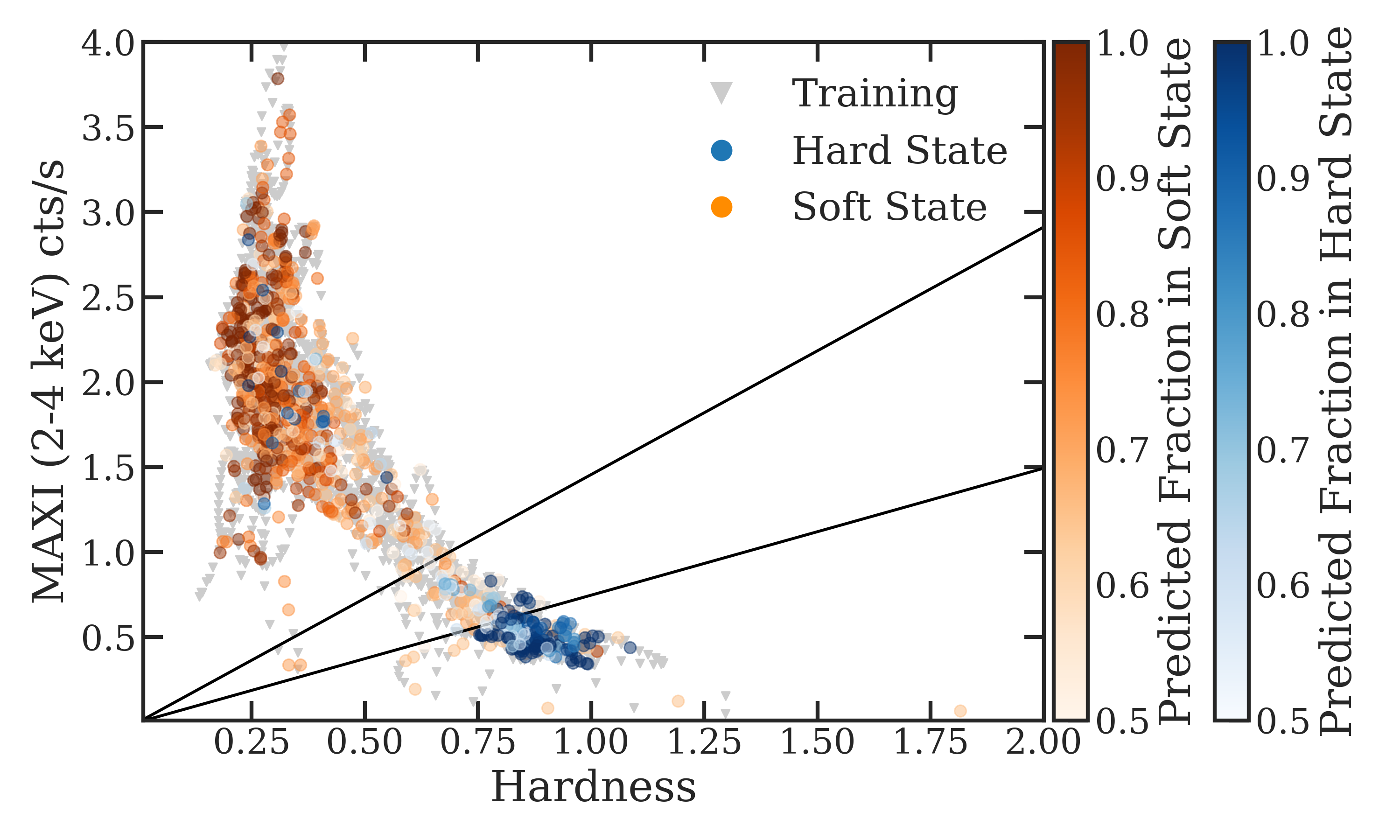}
	\includegraphics[width=\columnwidth]{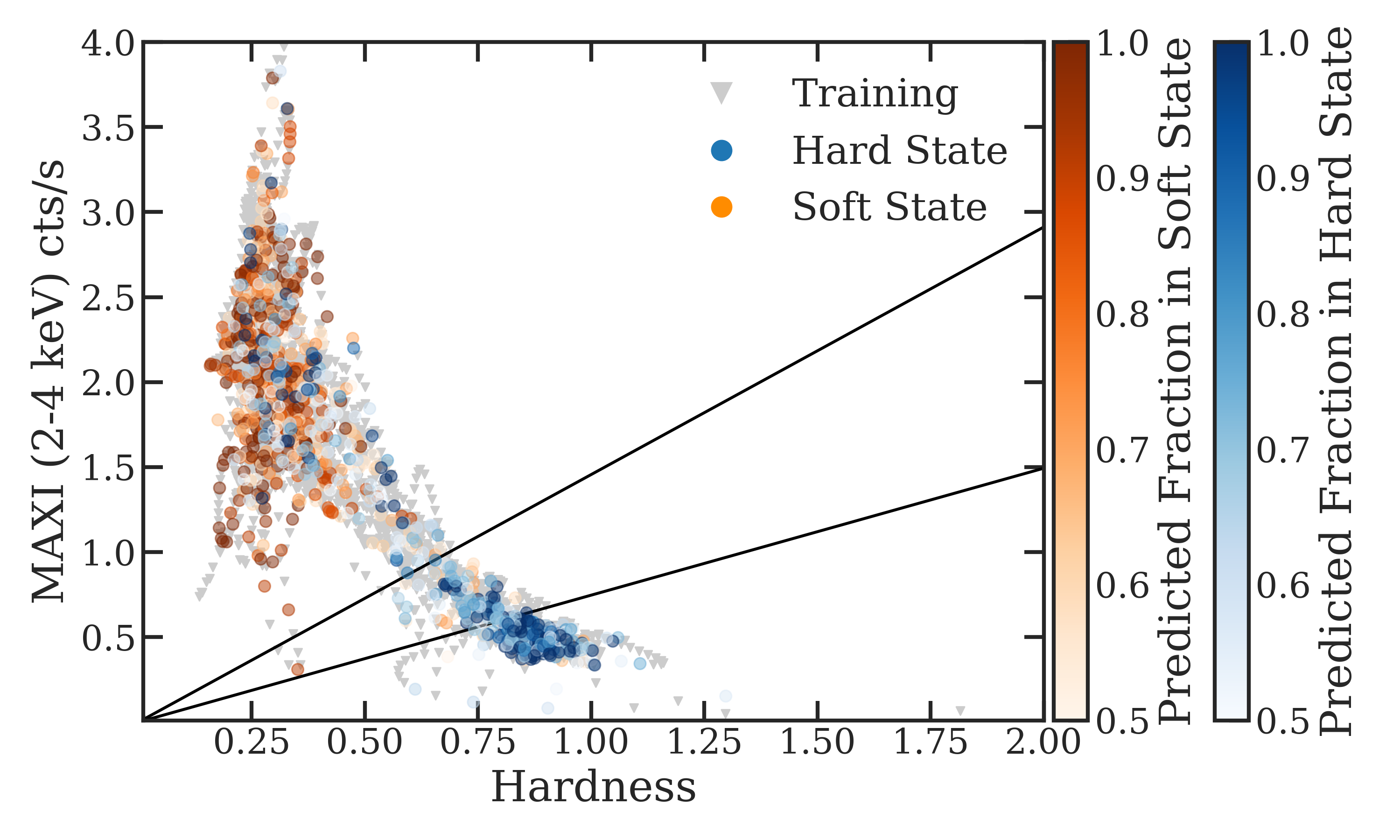}
    \caption{The same as Fig.~\ref{fig:HID_RXTE_predicted}, but for the MAXI HID (similar to Fig.~\ref{fig:HIDs}) colorized with the predicted values from the KNN regressor with a 250 d (top) or 100 d (bottom) RP window.}
    \label{fig:HID_MAXI_predicted}
\end{figure}

When analyzing the scores in Table \ref{tab:regressor_scores}, the ideal model will be one that has high accuracy at the smallest window size possible. Overall, we found that the accuracy scores lower as the RP window size decreases for all four types of models. The most severe drop in accuracy occurs between the 250-day and 100-day window sizes. For most models, the 2000, 1000, and 500-day windows have very similar AUC-ROC scores, with a small drop in accuracy occurring for the 250-day window. 

The KNN regressor has consistently higher AUC-ROC scores and lower RMSE. When comparing the different windows within the KNN regression model, the 2000, 1000, and 500-day windows have AUC-ROC scores equal to or greater than 99 per cent and the 250 day window has an AUC-ROC of 95 per cent for all spectral states. There is a similar pattern with the RMSE, where the 2000, 1000 and 500-day windows have RMSE scores of less than 5 per cent, and the 250-day window jumps above 5 per cent but below 10 per cent in each state. 

Ultimately, out of the 20 total models, we consider the 250-day window of the KNN regressor to satisfy our criteria of containing both a small RP window and high accuracy (above 95 per cent for all spectral states). The 250-day window KNN regressor model is visualized along with an example of a `poor' model (KNN regression with a 100-day window) in the HID for the RXTE-ASM data in Fig. \ref{fig:HID_RXTE_predicted}, and in the HID for MAXI in Fig. \ref{fig:HID_MAXI_predicted}. The hyperparameter options (as discussed in Sec. \ref{subsec:classifying via RP}) of the 250-day KNN regression model considered two neighboring points and applied weights that vary based on the distance to each neighbor.

\subsection{Feature Significance}

\begin{figure}
    \includegraphics[width=\columnwidth]{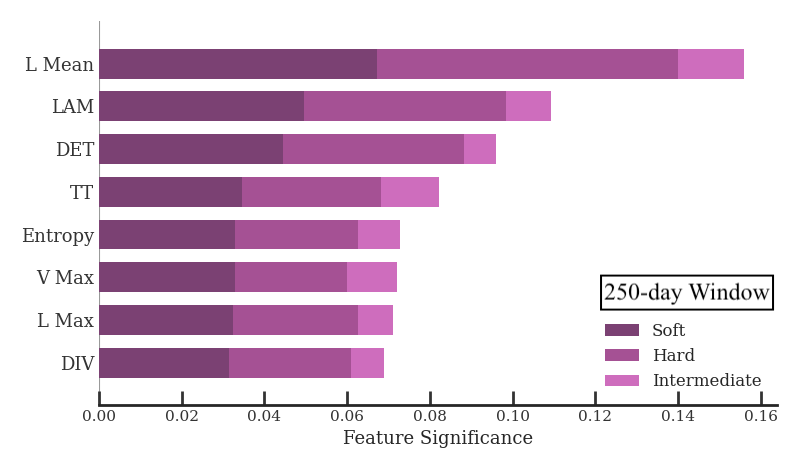}
	\includegraphics[width=\columnwidth]{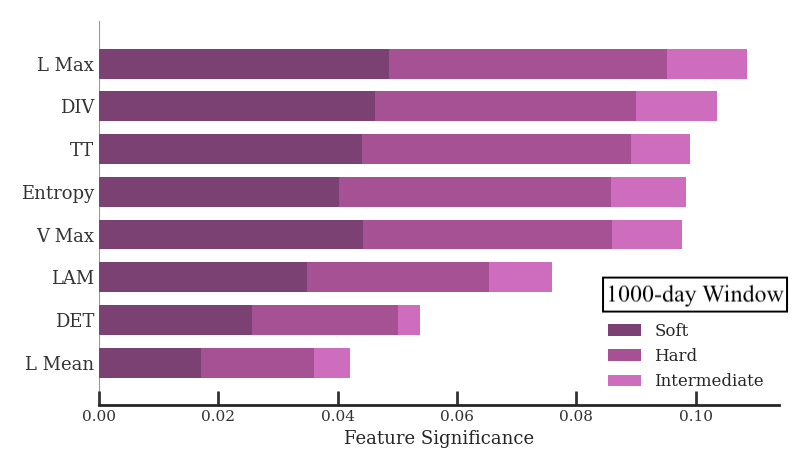}
    \caption{The feature significance of the eight RP statistics in the RF models for the (top) 250-d and (bottom) 1000-d RP window sizes using the $SHAP$ method \citep{Lundberg2017}.}
    \label{fig:shap}
\end{figure}

The correlations between RP statistics and spectral state, as depicted in Table~\ref{tab:t-test_fractional} of Sec.~\ref{subsec:correlations}, suggest that the most important features of an RP are the Shannon entropy and the average line lengths for determining accretion state for multiple RP window sizes. The feature importance of each statistic can be computed directly from the Random Forest regression model, which will also inform the relative importance of each RP feature to the ML models in determining the accretion state. We utilize the \textit{SHapley Additive ExPlanations} method ($SHAP$; \citealt{Lundberg2017}), which effectively assigns an importance value to each feature that represents the effect on the model prediction of including that feature. All five models differed in the order and relative strengths of the RP feature significances. Indeed, the most important feature resulted in $TT$, $L_{mean}$, $V_{max}$, $L_{max}$, and $V_{max}$, respectively, in order of increasing RP window size. Fig.~\ref{fig:shap} demonstrates this discrepancy in the feature significance for each of the eight RP statistics for the 250-d and 1000-d RP windows. We note how $L_{mean}$, $LAM$, and $DET$ were most important for the 250-day RP window, but were the three least important for the 1000-day RP window. Similar to Table~\ref{tab:t-test_fractional}, for example, $LAM$ increases importance for larger RP window sizes. The remaining three models were similar to the 1000-day model in that they all showed relatively comparable weights for most of the features. Only the 250-day model showed the top feature ($L_{mean}$) as significantly more important than the other features. 

Overall, our interpretation is that no features should be dropped from the ML models. Although the Shannon entropy and average line lengths were not consistently the top three features in the ML models for all window sizes, the results are still consistent with the observation that most recurrence features were significant for at least one RP window size in Table~\ref{tab:t-test_fractional}. This would be in line with the fact that the phase space geometries of real dynamical systems are rather complex, and thus no one statistic of a highly-dimensional space will be suitable for summarizing its complexity. In fact, a deep learning model that considers the recurrence matrix as a whole, rather than its summary statistics, would likely perform well for distinguishing subtle RP differences amongst data such as those from accreting systems.

\section{Conclusions}\label{sec:conclusions}

From the long-term monitoring of Cyg X-1 by the $RXTE$ All-Sky Monitor and the ongoing $MAXI$ instrument on the ISS, we construct hardness intensity diagrams to determine the spectral state of the source at each point in its light curve and compare to the variability characteristics described by the RP generated as a function of time.

First, we identify epochs, or extended periods of time, in the light curve for which the predominant spectral states or behavior of the source correlates to certain dynamical behavior inferred from the recurrence properties. These epochs suggest that there is deterministic behavior associated with an increased frequency of state changes in Cyg X-1. This observation echoes the findings from \cite{Sukova2016} that significant traces of nonlinear dynamics and determinism detected with recurrence analysis occur in systems that undergo the limit-cycle instability on short timescales. Here, on long timescales, we infer the preponderance of deterministic behavior likely correlates with a mechanism responsible for global accretion flow instabilities that is related to accretion state changes (e.g., \citealt{Cao2021}). In contrast, when the characteristic frequency of the turbulence (stochastic) is greater than the growth rate of nonlinear instabilities (deterministic), then we may observe, for example, something like a damped random walk in the observed emission \citep{Ross2017}. Indeed, the soft state of Cyg X-1 is dominated by red noise \citep{Zhou2022}. 

We also find that epochs with extended periods spent in the hard spectral state correspond to intermittent chaos or highly dimensional stochastic behavior, which may correspond to periods in which the source does not undergo a global instability in the accretion flow but nonetheless exhibits complex temporal variability. Given that Cyg X-1 exhibits persistent variability at radio wavelengths, with a significant detected radio jet in the LHS \citep{Stirling2001}, we posit that the increased dynamical complexity is due to corona-jet interactions and the introduction of a stronger disk component dampens this complexity. For example, it has been shown that the introduction of radiation (\citealt{Turner}) or self-gravity (\citealt{Riols}) of the accretion disk can dampen the strength of the magnetorotational instability and magnetic field of the accretion disk. Furthermore, it is expected that there are simultaneous, and interconnected, random and chaotic or periodic components in the light curves of accretion disk systems, operating on differing timescales, and the strength of each is dependent on the dominant sources of instability and the magnitude of stochastic fluctuations (\citealt{Voges1987, BoydSmale2004}). While the study of only Cyg X-1 is insufficient to make any distinct claims about the source of stochastic or deterministic behavior in accretion states generally, the distinct recurrence features in each spectral state of Cyg X-1 suggests intrinsic variability could be a distinguishing feature.

Second, using the classification into three spectral states for every observation in the Cyg X-1 light curve, we construct 20 different ML models (either a Random Forest or k-Nearest Neighbors regressor) to predict the spectral state based solely on the recurrence properties derived from the time-dependent RPs of the light curve. For all models explored, we find the accuracy of predicting the fractional amount of time spent in each spectral state for each RP remains high (above an approximately 80 per cent AUC-ROC score). The KNN regression model retains above a 95 per cent accuracy score in predicting all three spectral states for a window size as small as 250 days. That is, a light curve that contains 250 daily observations of a source is sufficient to predict whether it is predominantly in a soft, intermediate, or hard spectral state with up to 95 per cent accuracy. However, even 100 daily observations of a source would provide an adequate prediction of the dominant spectral state during that timeframe (an accuracy of 80 per cent). We, therefore, conclude that the RP provides a unique probe for determining the spectral state of an accreting source like Cyg X-1 based solely on its temporal variability characteristics. If such a model is successful in predicting the spectral states of other XRBs, then alerts of state changes can potentially be made with regular monitoring of accreting sources without the use of spectra.

Finally, we find that there are distinct differences in individual features of the RPs of Cyg X-1 between times when it is predominantly in the hard spectral state versus the soft spectral state. The recurrence features that correspond to information entropy, chaotic and laminar state intermittency, and recurrence memory are systematically higher in the hard spectral state than in the soft spectral state. This suggests that the accretion flow properties manifest distinct temporal variability characteristics. In particular, the LHS of Cyg X-1 (which also contains a radio jet detection; e.g., \citealt{Miller2012}) undergoes intermittency (periods of laminar, or time-invariant, behavior randomly interrupted by chaotic behavior) for longer periods of time relative to the HSS. This suggests that the decreased disk component and dominant coronal component in the hard state \citep{Pottschmidt2003, Basak2017} introduce more complex temporal variability, with an increased likelihood of determinism. In contrast, the HSS exhibits shorter recurrences, which could be due to more stochastic variability from the disk dominant state \citep{Zhou2022}, or potentially due to the decreased influence of the radio component in the soft state \citep{Zdziarski2020}.

Similar correlations between recurrences and variability or spectral states have been found. \cite{Sukova2016} has found that accretion states with specific kinds of QPOs will manifest with deterministic or chaotic dynamics detected by the RP that are distinct from stochastic variability evidenced in states without QPOs. Similarly, \cite{phillipson2020} found that a $Kepler$-monitored Seyfert 1 galaxy containing a low-frequency QPO exhibited more deterministic and nonlinear behavior compared to another Seyfert 1 galaxy without a QPO. And, \cite{Phillipson2023} has found that Type 1 and Type 2 AGN exhibit distinct RPs, as do radio-quiet and radio-loud AGN, which supports models of AGN with differing accretion states akin to XRBs. In particular, the common thread through all of these studies may be the role of the radio jet, which could induce deterministic, nonlinear, or chaotic modulations in the emission. At a minimum, the distinct classifications based on recurrence properties that exist in the different spectral states of Cyg X-1, variability states of microquasars (as in \citealt{Sukova2016}), and classes of AGN suggest that the dynamics of the accretion flow are imprinted in the light curves of accreting sources. These results suggest that novel time series analysis approaches combined with machine learning can potentially be leveraged for the study of accretion without spectra. 

The next step in our study is to extend the analysis to other state-changing XRBs to determine whether the distinct recurrence features persist for other sources in the different spectral states in the same manner as Cyg X-1, or if there are further dependencies on other characteristics in the system, such as the companion star mass or mode of accretion (e.g., Roche-lobe overflow versus stellar wind accretion). The ultimate goal is to leverage information from the RP for the classification of ensemble studies of accreting sources, or for the discovery of new accreting sources in large time domain surveys.

\section*{Acknowledgements}

The authors thank Joey Neilsen and Eric Bellm for helpful discussions regarding the manuscript and analysis results. The authors also thank the anonymous referee for comments that improved the manuscript. R.A.P. acknowledges support for this work provided by the National Science Foundation (NSF) MPS-Ascend Postdoctoral Research Fellowship under Grant No. 2138155. E.M.B acknowledges support for this work through the Washington NASA Space Grant Consortium Summer Undergraduate Research Program. R.A.P. and E.M.B. both acknowledge support for this work through a gift of the Washington Research Foundation to the University of Washington eScience Institute and from the NSF Astronomy and Astrophysics Research Grants (AAG) Program under Grant No. 1812779.

\section*{Data Availability}

This research made use of data retrieved from the publicly available repository of the Rossi X-ray Timing Explorer All-Sky Monitor at the NASA High-Energy Astrophysics Science Archives Research Center (https://heasarc.gsfc.nasa.gov/docs/xte/xhp\_archive.html) and from the public archive provided by the RIKEN, JAXA, and MAXI team (http://maxi.riken.jp/). The code used to analyze the data includes the publicly available packages PyUnicorn (\citealt{Donges2015}; available at http://www.pik-potsdam.de/$\sim$donges/pyunicorn) for the production of the Legendre coordinate phase space embeddings; the Scikit-Learn modules from Python for the machine learning models; and methods from \cite{Schinkel2009} and the Matlab \textit{CRP-Toolbox} (https://tocsy.pik-potsdam.de/CRPtoolbox/) translated into Python for the RPs, RP features, and the confidence intervals of RP features. The combination of these packages for use in the analyses in this paper was facilitated by scripts written in Python and will be shared on reasonable request to the corresponding author.



\bibliographystyle{mnras}
\bibliography{References} 


\appendix

\section{Comparisons to the Intermediate State}\label{sec:pvalues}

Here we include the resulting p-values from the t-test comparing the mean RP features between the soft/hard and intermediate spectral states in Tables~\ref{tab:t-test_appendixA} and~\ref{tab:t-test_appendixB}, respectively. These are the same as that described in Sec.~\ref{subsec:correlations} and in Table~\ref{tab:t-test_fractional}. Here we find no significant differences between the intermediate state and the other two states.

\begin{table}
\caption {The same as Table 4: The p-values obtained from Welch's $t$-Test comparing the means of RP features between the intermediate and soft spectral states.} \label{tab:title}
\begin{tabular}{ |c|c|ccccc|} 
\hline
 & \multicolumn{5}{|c|}{Window Size (days)} \\
\hline
Feature       &  2000 & 1000 & 500  & 250  & 100\\
\hline
Determinism:  & 0.48  & 0.16 & 0.15 & 0.23 & 0.61\\
Laminarity:   & 0.49  & 0.16 & 0.08 & 0.34 & 0.63\\
Entropy:      & 0.50  & 0.10 & 0.14 & 0.19 & 0.63\\
L Mean:       & 0.72  & 0.18 & 0.19 & 0.17 & 0.55\\
TT:           & 0.65  & 0.16 & 0.14 & 0.37 & 0.61\\
L Max:        & 0.43  & 0.11 & 0.16 & 0.54 & 0.63\\
V Max:        & 0.56  & 0.42 & 0.15 & 0.56 & 0.61\\
Divergence:   & 0.43  & 0.10 & 0.21 & 0.48 & 0.64\\
\hline

\hline
\end{tabular}\label{tab:t-test_appendixA}
\end{table}

\begin{table}
\caption {The same as Table 4: The p-values obtained from Welch's $t$-Test comparing the means of RP features between the intermediate and hard spectral states.} \label{tab:title}
\begin{tabular}{ |c|c|ccccc|} 
\hline
 & \multicolumn{5}{|c|}{Window Size (days)} \\
\hline
Feature       &  2000 & 1000 & 500  & 250  & 100\\
\hline
Determinism:  & 0.52  & 0.66 & 0.66 & 0.63 & 0.66\\
Laminarity:   & 0.71  & 0.61 & 0.61 & 0.64 & 0.65\\
Entropy:      & 0.10  & 0.72 & 0.71 & 0.66 & 0.58\\
L Mean:       & 0.11  & 0.71 & 0.70 & 0.66 & 0.66\\
TT:           & 0.10  & 0.71 & 0.71 & 0.69 & 0.58\\
L Max:        & 0.72  & 0.39 & 0.50 & 0.64 & 0.60\\
V Max:        & 0.42  & 0.71 & 0.67 & 0.65 & 0.44\\
Divergence:   & 0.63  & 0.66 & 0.63 & 0.59 & 0.57\\
\hline

\hline
\end{tabular}\label{tab:t-test_appendixB}
\end{table}

\section{Recurrence Plots of Dynamical Systems}\label{sec:app_rps}

\begin{figure*}
	\includegraphics[width=\textwidth]{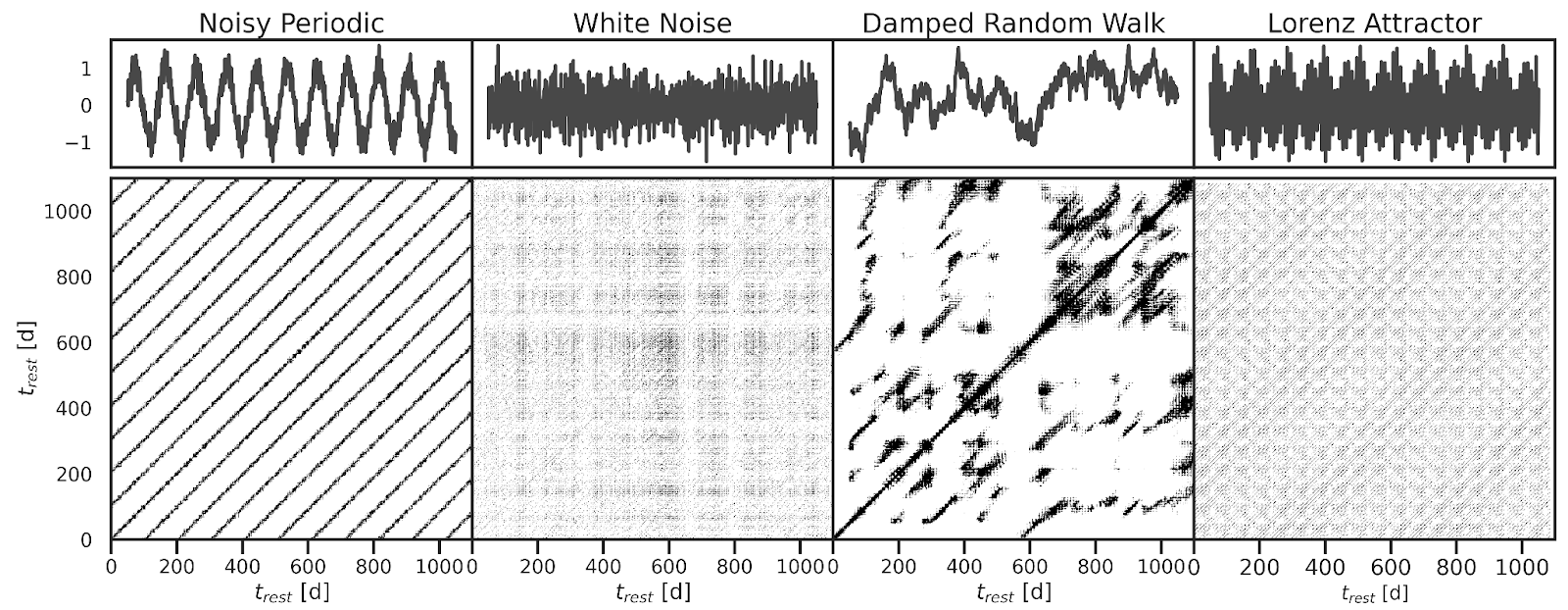}
    \caption{Examples of four canonical time series (top row) and their corresponding RPs (bottom row). From left to right: a noisy periodic signal in which the diagonal lines are separated by the period; white noise leading to randomly distributed points; a first-order autoregressive process (damped random walk) leading to diagonally oriented structures that vary in scale (representative of correlations); and the Lorenz attractor operating in the chaotic regime, leading to recurrent patterns throughout the RP including chopped up diagonal lines.}
    \label{fig:rps_examples}
\end{figure*}

Fig.~\ref{fig:rps_examples} provides four example time series and their corresponding RPs. The evenly spaced diagonal lines (parallel to the line of identity) are formed when the trajectory appears periodically at the same place in the phase space on more than one specific time, known as a recurrence. The lengths of these lines highlight the time duration of the recurrence and can be quantified by Recurrence Quantification Analysis (RQA; \citealt{Webber1994}), which utilizes various statistical distributions of line lengths to describe structure. Diagonal lines are typically a sign of deterministic behavior, as you can see in the periodic example (left panel) of the RP in Fig.~\ref{fig:rps_examples}. Horizontal and vertical structures are times when the trajectory does not vary strongly, typically a sign of laminarity or time invariance. Chaotic variability results in many chopped-up diagonal lines, as evidenced by the chaotic Lorenz attractor in the right panel of Fig.~\ref{fig:rps_examples}. Uncorrelated noise will lead to randomly distributed points. 

\subsubsection{Phase Space: The Method of Delays}\label{sec:phase_space}

For a given scalar series, $\vec{x}(t)$, the time delay method \citep{Takens1981} maps the observations to time-delay embedded vectors $\vec{y}(t)$ by creating an $n$ vector map:
\begin{equation}
\begin{split}
    x(t) \rightarrow y(t) & = (y_1(t), y_2(t), ..., y_m(t)) \\
    y_j(t) & = x(t - \tau_j), \, j = 1, 2, ..., m,
\end{split}\label{eq:delays}
\end{equation}
where $m$ is the dimension of the embedded vector and the time delay is defined as $\tau = k\Delta t$, where $\tau$ is an integer ($k$) multiple of the cadence of the light curve, $\Delta t$.

\begin{figure}
    \includegraphics[width=\columnwidth]{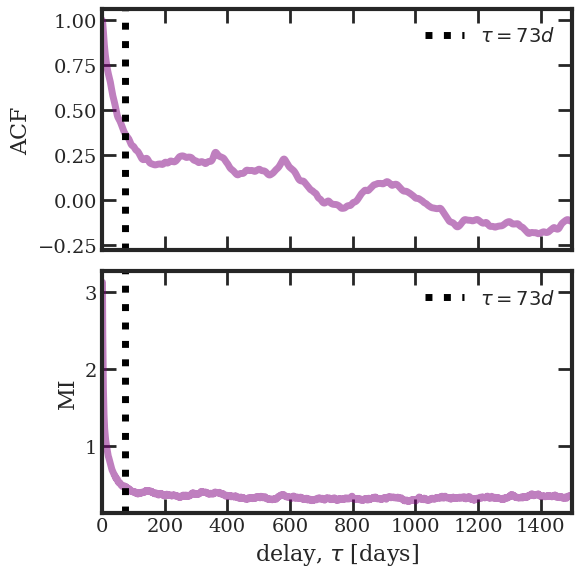}
    \caption{(top) The autocorrelation function and (bottom) mutual information for Cyg X-1 for a delay of up to 1500 days. The vertical dashed line corresponds to 73 days, which is the autocorrelation time derived from the ACF.}
    \label{fig:ACF_MI}
\end{figure}

In order to effectively reconstruct the underlying attractor that generates a time series, the time-delayed vectors should contain components that are uncorrelated. In other words, the time delay, $\tau$, should be larger than the correlation time in the time series. The autocorrelation time derived from the autocorrelation function (ACF) is often used, though the first minimum in the mutual information (MI; \citealt{Fraser1986}) is also appropriate for time series with possible nonlinear correlations. The ACF and MI for Cyg X-1 are shown in Fig.~\ref{fig:ACF_MI}.

\begin{figure}
    \includegraphics[width=\columnwidth]{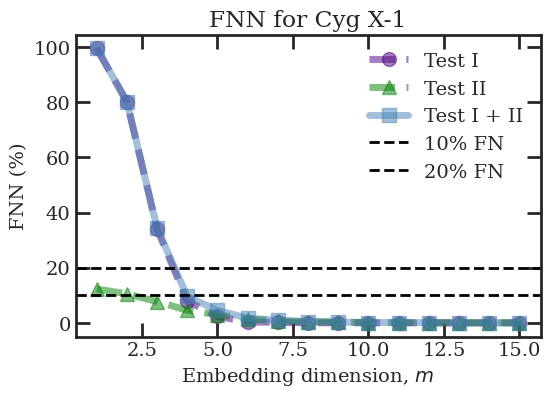}
    \caption{The FNN method from \protect\cite{Kennel1992}, showing the two statistical tests for determining false neighbors in dashed lines of purple and green. The third, solid blue line is merely the sum of the two tests. The horizontal dashed lines represent 10 per cent and 20 per cent false neighbors for reference. Cyg X-1 reaches less than 1 per cent false neighbors for an embedding dimension of 8 for all three statistical tests.}
    \label{fig:FNN}
\end{figure}

The false nearest neighbors (FNN) method \citep{Kennel1992} is traditionally used for selecting an appropriate embedding dimension, $m$. This involves iteratively increasing the dimension until the number of false neighbors (neighboring points in phase space that diverge from each other when the dimension is increased) is minimized. The FNN for up to 15 embedding dimensions of Cyg X-1 is presented in Fig.~\ref{fig:FNN} using the two statistical tests from \cite{Kennel1992}. For the first test, the threshold that dictates whether these neighbors are `false' is set by a fixed value. For the second text, the relative distance between each pair of points identified as neighbors is compared between dimensions $m$ and $m+1$. If the ratio of the relative distances to the standard deviation of the input time series is greater than a fixed threshold, then these neighbors are considered `false.'

For the case of Cyg X-1, we select a time delay from the autocorrelation time and the embedding dimension from the FNN algorithm, which results in $\tau = 73$ (or 73 days, given daily monitoring) and $m = 8$. 

\subsubsection{Phase Space: Legendre Coordinates}

The phase space can also be reconstructed using orthogonal polynomial filters, called `Legendre coordinates' \citep{Gibson1992}.
To summarize the method from \cite{Gibson1992}, the $j$th-order derivative of a time series can be estimated by a discrete linear filter,
\begin{equation}\label{eq:filter}
    w_j(t) = \sum_{n=-p}^p r_{j,p}(n) \cdot x(t+nr),
\end{equation}
where the time series $x(t)$ is the input, $w_j(t)$ is the output, and $r_{j,p}(n)$ is an appropriate discrete convolution kernel, parameterized by the (arbitrary) choice of $p$ and the order of the desired derivative, $j$. By expanding $x(t)$ in a Taylor series and enforcing orthogonality constraints, \cite{Gibson1992} demonstrates that the output, $w_j(t)$, is proportional to the $j$th derivative of the time series. In the limit that $p$ approaches infinity, the kernels $r_{j,p}(n)$ reduce to the Legendre polynomials; \cite{Gibson1992} thus refers to them as discrete Legendre polynomials.

The discrete Legendre polynomials form an orthonormal basis, with basis vectors $\vec{r}_j$. As such, we can project a delay vector (e.g., $y(t)$ from Eq.~\ref{eq:delays}) onto the discrete Legendre polynomial basis by Eq.~\ref{eq:filter}:
\begin{equation}
    w_j(t) = \vec{y}(t) \cdot \vec{r}_j,
\end{equation}
which \cite{Gibson1992} thus labels a `Legendre coordinate' that is proportional to a derivative of the original time series (i.e., $w_j(t) \propto x^j(t)$). We can therefore reconstruct the phase space trajectory using Legendre coordinates that are proportional to the derivatives of the system. This becomes particularly useful in the context of irregularly spaced intervals and for time series for which the derivatives and underlying equations of motion are unknown.

\subsubsection{Recurrence Quantification Analysis}\label{sec:rqa}

Following the notation of \cite{marwan2007}, we define RQA statistics --- or RP features, as we refer to them throughout the paper --- that can be used to summarize the overall properties of an RP, namely, the Recurrence Rate, Determinism, Laminarity, average vertical/horizontal and diagonal line lengths, longest diagonal and vertical line lengths, Divergence, and Shannon Entropy. As a collective, these features summarize the line structures and texture of the RP for a given threshold, $\epsilon$.

The recurrence rate (RR) is the probability that a state returns to within an $\epsilon$-neighborhood in phase space:
\begin{equation}\label{eq:rr}
RR(\epsilon )=\frac{1}{N^2}\sum ^{N}_{i,j=1}\mathbf{R}_{i,j}(\epsilon ),
\end{equation}
where N is the length of the time series and $\mathbf{R(\epsilon)}$ is the recurrence matrix for a given threshold, $\epsilon$. The RR describes the density of recurrence points in the RP.

Determinism ($DET$) is the percentage of points that form a diagonal line compared to isolated points in the RP: 
\begin{equation}\label{eq:det}
DET=\frac{\sum^{N}_{l=l_{min}} lP(l) }{\sum^{N}_{l=l} lP(1)},
\end{equation}
where $l_{min}$ is the minimum length of the diagonal lines found in the RP (typically set to 2) and $P(l)$ is a histogram of all the diagonal line lengths, $l$. The higher $DET$, the more recurrence points are part of diagonal structures and the less randomness is present in the system.

Similarly, Laminarity ($LAM$) is the ratio of the number of recurrence points that form vertical structures, $P(v)$, to the total number of recurrence points in an RP,
\begin{equation}\label{eq:lam}
LAM = \frac{\sum_{v=v_{\min}}^N v P(v)}{\sum_{v=1}^N v P(v)},
\end{equation}
where $v_{min}$ is the minimum length of the vertical lines found in the RP (typically set to 2) and $P(v)$ is a histogram of all the vertical line lengths, $v$. Thus, $LAM$ is analogous to $DET$ and measures the frequency of laminar (or time-invariant) states in the system relative to randomness.

The Trapping Time ($TT$) represents the amount of time the trajectory remains in one state, or acute region of phase space, where
\begin{equation}\label{eq:TT}
TT = \frac{\sum_{v=v_{\min}}^N v P(v)}{\sum_{v=v_{\min}}^N  P(v)},
\end{equation}
representing the average length of a vertical line. $TT$ is also considered to be the average time in which fluctuations occur in an impulse-response system \citep{phillipson2020}.

The longest length of diagonal lines ($L_{max}$) and vertical lines ($V_{max}$) of the RP are analogously defined: 
\begin{equation}\label{eqn:max}
\begin{split}
L_{\max} &= \max\left( \{l_{i}; \ i=1,\ldots, N_{l}\}\right),\\
V_{\max} &= \max\left( \{v_{i}; \ i=1,\ldots, N_{v}\}\right),
\end{split}
\end{equation}
with $N_l=\sum_{l\ge l_{\min}}P(l)$ and $N_v=\sum_{v\ge v_{\min}}P(v)$ as the total number of diagonal lines and vertical lines, respectively, in the RP. 

Divergence is known as the inverse of $L_{max}$, defined as
\begin{equation}\label{eq:div}
DIV = \frac{1}{L_{\max}},
\end{equation}
where the shorter the divergence between trajectories in phase space, the smaller the $DIV$ feature, which creates longer diagonal lines in the RP. The inverse relationship between $DIV$ and $L_{max}$ is explained using the sum of the positive Lyapunov exponents of the system \citep{trulla1996}.

Finally, the Shannon Entropy is defined as:
\begin{equation}\label{eq:entr}
ENTR = -\sum_{l=l_{\min}}^N p(l) \ln p(l),
\end{equation}
where $p(l)=P(l)/N_l$ is the probability that a diagonal line is of length $l$ in the RP. The Shannon entropy is also considered a measure of information, where an increase in entropy corresponds to greater uncertainty in the time series distribution.

\section{Constructing the RP of Cygnus X-1}

The optimal threshold for performing recurrence analysis corresponds to at least 10 per cent of the maximum diameter of the phase space (\citealt{Schinkel2008}) and exceeding 5 times the standard deviation of the observational noise (\citealt{Thiel2002}), but not exceeding the maximum size of the phase space \citep{Thiel2002}. For example, for a 3-dimensional time series that has an amplitude range of unity and noise with a standard deviation of 0.02, the threshold choice should be at least 1.7 and greater than 5$\sigma_{obs} = 1$.

Given that the Legendre coordinates approximate the derivatives of the one-dimensional time series, this method has an advantage over the time delay method in its application to shorter time series. For example, embedding the full light curve of Cyg X-1 into phase space using the time delay method uses a delay of 73 days and a dimension of 8, which corresponds to an embedding window of 584 days. This means we would not be able to explore window sizes in the windowed RP shorter than the embedding window. Given that we seek to explore window sizes as small as possible, much less than 500 days, in order to isolate individual spectral states as much as possible, we use the Legendre coordinates method. The drawback is that utilizing the derivative of the time series results in greater noise in the phase space embedding. However, if we are able to distinguish spectral states using RQA features despite this added noise (in contrast to the noise suppression that naturally occurs in the time delay method), then we can say that features are robust against noise contamination for classification and regression purposes.

\section{AUC-ROC Score}\label{sec:AUC_ROC_Score}

For a binary classification model for which labels are either `positive' or `negative,' there are four possible outcomes: true positives, false positives, true negatives, and false negatives. The true positive rate (TPR) is defined as the fraction of correctly predicted positives (true positives) relative to the total number of actual positives in the dataset. The false positive rate (FPR) is defined as the fraction of points incorrectly predicted as positive (false positives) relative to the total number of actual negatives in the dataset. 

The $ROC$ curve is generated by plotting the TPR as a function of FPR for each decision boundary between the binary classification. That is, for some decision boundary, $d$, we consider the positives as those labels with a probability of being positive greater than $d$ and the negatives as those labels with a probability of being positive less than $d$. The TPR and FPR are calculated for each $d$ and the ROC curve therefore runs from zero to one. The resulting ROC curve gives an indication of the trade-offs between acquiring true positive results versus false positive results. A ROC curve that follows a slope of one corresponds to a classifier that is no better than the flip of a coin. The higher the TPR is relative to FPR for a wide range of decision boundaries, the more accurate the classifier is in both predicting the true labels of the data and limiting the number of false positives.

A means to quantify the quality of the ROC curve in a single number is to calculate the area under the ROC curve (typically done by the trapezoid rule), referred to as the AUC. The resulting area is called the AUC-ROC score. A perfect model would result in an AUC-ROC score of unity.

For the case of a regression model, we must consult an alternative definition of the true and false positive rates, since the predicted and real labels of the data are continuous variables (in this case, the fraction of observations that are in each of the spectral states for a given sub-RP window). The TPR is also known as the probability of detection and the FPR is also known as the probability of false alarm. That is, for the decision boundary, $d$, the probability distribution for the positive label, $f_p(x)$, and the probability distribution for the negative label, $f_n(x)$, we define 
\begin{equation}
    \begin{split}
        TPR &= \int_{d}^{\infty} f_p(x) \,dx, \\
        FPR &= \int_{d}^{\infty} f_n(x) \,dx.
    \end{split}
\end{equation}

Thus, if we know the probability distributions for both detection and false alarm from the dataset, then the ROC curve becomes equivalent to the cumulative distribution function.

For a finite dataset, the cumulative distribution can be estimated by determining the probability that a classifier will rank a randomly chosen positive instance from the dataset higher than a randomly chosen negative one. To implement, we consider each pair of predictions, $y_i$ and $y_j$, in the dataset, where $i = 1, ..., N$ and $j = 1, ..., N$ (for $i \ne j$). Then, the AUC-ROC score becomes the sum of instances where $y_i > y_j$ relative to all possible pairs. In Sec.~\ref{subsec:assessment and accuacy}, we consider each spectral class prediction separately, where the spectral state of interest is considered the `positive' label, and the other two spectral states as the `negative' label. 
\label{lastpage}
\end{document}